\documentclass[english,journal,12pt,onecolumn,draftclsnofoot]{IEEEtran}
\usepackage[latin9]{inputenc}
\usepackage{color}
\usepackage{array}
\usepackage{booktabs}
\usepackage{mathrsfs}
\usepackage{multirow}
\usepackage{amsmath}
\usepackage{amssymb}
\usepackage{graphicx}

\makeatletter

\providecommand{\tabularnewline}{\\}

\usepackage[center]{caption}
\usepackage[font=footnotesize]{caption}
\usepackage{flushend}

\@ifundefined{showcaptionsetup}{}{%
 \PassOptionsToPackage{caption=false}{subfig}}
\usepackage{subfig}
\makeatother

\usepackage{babel}
\begin{document}
\title{\vspace{-2mm}
Hybrid-Layers Neural Network Architectures for Modeling the Self-Interference
in Full-Duplex Systems}
\author{Mohamed Elsayed, \textit{Student Member, IEEE}, Ahmad A. Aziz El-Banna,
\textit{Member, IEEE}, Octavia A. Dobre, \textit{Fellow, IEEE}, Wanyi
Shiu, and Peiwei Wang\vspace{-9mm}
 \thanks{M. Elsayed, A. A. A. El-Banna, and O. A. Dobre are with the Faculty
of Engineering and Applied Science, Memorial University, St. John\textquoteright s,
NL A1B 3X5, Canada (e-mail: \{memselim, aaelbanna, odobre\}@mun.ca).} \thanks{A. A. A. El-Banna is also with the Faculty of Engineering at Shoubra,
Benha University, Banha, Egypt.} \thanks{W. Shiu and P. Wang are with Huawei Canada Research Centre, Huawei
Technologies Canada Co., Ltd., Ottawa, ON K2K 3J1, Canada (e-mail:
\{wanyi.shiu, peiwei.wang\}@huawei.com).}}
\maketitle
\begin{abstract}
Full-duplex (FD) systems have been introduced to provide high data
rates for beyond fifth-generation wireless networks through simultaneous
transmission of information over the same frequency resources. However,
the operation of FD systems is practically limited by the self-interference
(SI), and efficient SI cancelers are sought to make the FD systems
realizable. Typically, polynomial-based cancelers are employed to
mitigate the SI; nevertheless, they suffer from high complexity. This
article proposes two novel hybrid-layers neural network (NN) architectures
to cancel the SI with low complexity. The first architecture is referred
to as hybrid-convolutional recurrent NN (HCRNN), whereas the second
is termed as hybrid-convolutional recurrent dense NN (HCRDNN). In
contrast to the state-of-the-art NNs that employ dense or recurrent
layers for SI modeling, the proposed NNs exploit, in a novel manner,
a combination of different hidden layers (e.g., convolutional, recurrent,
and/or dense) in order to model the SI with lower computational complexity
than the polynomial and the state-of-the-art NN-based cancelers. The
key idea behind using hybrid layers is to build an NN model, which
makes use of the characteristics of the different layers employed
in its architecture. More specifically, in the HCRNN, a convolutional
layer is employed to extract the input data features using a reduced
network scale. Moreover, a recurrent layer is then applied to assist
in learning the temporal behavior of the input signal from the localized
feature map of the convolutional layer. In the HCRDNN, an additional
dense layer is exploited to add another degree of freedom for adapting
the NN settings in order to achieve the best compromise between the
cancellation performance and computational complexity. The complexity
analysis of the proposed NN architectures is provided, and the optimum
settings for their training are selected. The simulation results demonstrate
that the proposed HCRNN and HCRDNN-based cancelers attain the same
cancellation of the polynomial and the state-of-the-art NN-based cancelers
with an astounding computational complexity reduction. Furthermore,
the proposed cancelers show high design flexibility for hardware implementation,
depending on the system demands. 
\end{abstract}


\vspace{-2mm}

\section{Introduction }

\def\figurename{Fig.}
\def\tablename{TABLE}

\IEEEPARstart{R}{ecently}, the evolution of Internet-of-Everything,
supporting massive connectivity among billions of users and billions
of devices, has imposed a radical shift towards the next generation
of wireless networks, such as beyond the fifth-generation (B5G) {[}\ref{Toward 6G networks}{]}.
These modern networks aim to provide high reliability, low latency,
and high data rates, in the order of tens Gbits/s, to enable extended
reality applications, live multimedia streaming, and autonomous systems
in smart cities and factories, such as drone swarms, cars, and robotics
{[}\ref{A vision of 6G wireless systems}{]}, {[}\ref{On the spectral and energy efficiencies of full-duplex cell-free}{]}.
As such, to cater to this novel breed of applications and support
such a plethora of services, the next generation of wireless systems
should be inherently tailored to simultaneously deliver higher data
rates with lower communication delays for both uplink and downlink. 

In this regard, full-duplex (FD) has emerged as one of the key enabling
technologies for B5G wireless networks by providing high data rates
through simultaneous transmission of information over the same frequency
resources {[}\ref{A.-Sabharwal,-P.}{]}-{[}\ref{K.-E.-Kolodziej,}{]}.
Efficient exploitation of the resources enables the FD systems to
meet the high quality-of-service requirements in terms of spectral
efficiency, which represents a major factor in designing B5G wireless
networks. On the other hand, the main challenge in implementing the
FD systems is the self-interference (SI), which comes out from the
transmitter of the same device on its own receiver {[}\ref{Experiment-driven characterization of full-duplex}{]}.
This undesirable interference significantly hinders the proliferation
of FD systems in the next generation of wireless networks {[}\ref{A.-Sabharwal,-P.}{]}-{[}\ref{K.-E.-Kolodziej,}{]}.

Over the past decade, a flurry of research interest has been directed
for canceling the interference in FD systems to make them realizable
{[}\ref{K.-E.-Kolodziej,}{]}, {[}\ref{Experiment-driven characterization of full-duplex}{]}.
Typically, canceling the SI can be implemented in analog radio frequency
(RF) and/or digital domains to bring the SI signal\textquoteright s
power down to the receiver\textquoteright s noise level. The analog
RF suppression is implemented at the very first stage of the receiver
chain to refrain the SI signal from saturating the analog components
of the receiver, such as the low-noise amplifier (LNA), variable-gain
amplifier (VGA), and analog-to-digital converter (ADC) {[}\ref{K.-E.-Kolodziej,}{]}.
In particular, the analog RF cancellation can be classified into passive
and active cancellations {[}\ref{Passive self-interference suppression}{]},
{[}\ref{On self-interference suppression methods}{]}. Passive cancellation
is implemented using techniques such as antenna separation {[}\ref{Full-duplex wireless communications using off-the-shelf}{]},
circulators {[}\ref{Simultaneous transmission and reception:}{]},
polarized antennas {[}\ref{Analog/RF solutions enabling compact full-duplex radios}{]},
and balanced hybrid-junction networks {[}\ref{Optimum single antenna full duplex using hybrid junctions}{]}.
On the other side, the active suppression is performed using analog
circuits, which generate a copy of the SI signal in order to be subtracted
from the original SI signal at the receiver chain {[}\ref{On self-interference suppression methods}{]}.
In general, the analog suppression techniques are insufficient to
entirely remove the SI at the receiver side, and a non-negligible
residual SI still exists after the analog cancellation process. Hence,
digital cancellation approaches are utilized in order to mitigate
the residual interference {[}\ref{All-digital self-interference cancellation technique}{]}. 

Digital domain cancellation uses the same notion of active suppression
where a processed copy of the baseband transmitted signal is subtracted
from the residual SI signal, but in the digital domain {[}\ref{All-digital self-interference cancellation technique}{]}.
In principle, the digital cancelers could effectively eliminate this
SI signal since it stems from a transmit signal that is obviously
known to the receiver. However, this is not the case in practice,
as the SI signal is significantly distorted by the SI coupling channel
and the impairments of the transceiver components, such as non-linear
distortion of the power amplifier (PA), in-phase and quadrature-phase
(IQ) imbalance of the mixer, phase noise of imperfect transceiver's
oscillators, and digital-to-analog converter (DAC) and ADC's quantization
noise {[}\ref{D.-Korpi,-L.}{]}.

In order to efficiently cancel the SI in the digital domain, the digital
cancelers should properly model the distortion incorporated into the
input signal due to the imperfection of the hardware and the SI channel.
Generally, modeling the transceiver's impairments is based on the
polynomial approximation of the SI signal at the receiver side {[}\ref{All-digital self-interference cancellation technique}{]}.
The polynomial-based models have excellent modeling capabilities to
mimic the SI signal; however, they suffer from high complexity {[}\ref{D.-Korpi,-L.}{]}.
Accordingly, low-complexity modeling approaches are sought for approximating
the SI signal in FD systems. 

Applying neural networks (NNs) and deep learning has gained significant
momentum in the field of signal processing and wireless communications
in the last few years {[}\ref{An introduction to deep learning for the physical layer}{]}-{[}\ref{Nonlinear interference mitigation via deep neural networks}{]}.
NNs have been recently employed to replace the model-based approaches
in numerous communication areas in order to approximate the non-linearities
with good performance and low implementation complexity. For instance,
NNs have brought breakthroughs in signal detection {[}\ref{An introduction to deep learning for the physical layer}{]},
signal classification {[}\ref{Joint blind identification of the number of transmit antennas and MIMO schemes}{]},
channel estimation {[}\ref{Power of deep learning for channel estimation and signal detection}{]},
{[}\ref{Deep learning-based channel estimation for}{]}, channel equalization
{[}\ref{Deep learning-based channel estimation and equalization scheme}{]},
channel coding {[}\ref{On deep learning-based channel decoding}{]},
PA modeling {[}\ref{F.-Mkadem,-M.}{]}-{[}\ref{Paper of (11) }{]},
digital pre-distortion {[}\ref{F.-Mkadem-and S. Boumaiza}{]}-{[}\ref{Deep neural network-based digital predistorter}{]},
and non-linearity compensation in optical fiber systems {[}\ref{Nonlinear interference mitigation via deep neural networks}{]}.

In addition, there has been a surge of interest in applying NNs for
SI cancellation in FD systems {[}\ref{A.-Balatsoukas-Stimming,-"Non-li}{]}-{[}\ref{Our paper }{]}.
More specifically, the first attempt of using NNs for canceling the
SI has been reported in {[}\ref{A.-Balatsoukas-Stimming,-"Non-li}{]},
where a real-valued time delay NN (RV-TDNN)\footnote{\label{fn:footnote1}We note that the RV feed-forward NN (RV-FFNN)
in {[}\ref{A.-Balatsoukas-Stimming,-"Non-li}{]} has a similar structure
to the RV-TDNN in {[}\ref{F.-Mkadem,-M.}{]} since both employ the
input signal's buffered samples at the input layer. Henceforth, we
will use RV-TDNN instead of RV-FFNN for accurate referring.} is introduced to model the SI signal with computational complexity
lower than the polynomial-based canceler. In {[}\ref{A.-T.-Kristensen,}{]},
a recurrent NN (RNN) and a complex-valued TDNN (CV-TDNN) have been
investigated for SI mitigation; it is shown that the CV-TDNN has excellent
modeling capabilities to approximate the SI with lower computational
complexity than the polynomial and RNN-based cancelers. In {[}\ref{Design and implementation of a neural network aided self-interference}{]},
{[}\ref{Hardware implementation of neural self-interference cancellation}{]},
the hardware design of the polynomial and NN-based cancelers introduced
in {[}\ref{A.-Balatsoukas-Stimming,-"Non-li}{]} have been provided.
Furthermore, in {[}\ref{Our paper }{]}, the ladder-wise grid structure
(LWGS) and moving-window grid structures (MWGS), two low-complexity
NN models, have been introduced for SI cancellation. It is demonstrated
that the LWGS and MWGS attain a similar cancellation performance to
the polynomial and CV-TDNN-based cancelers with a significant complexity
reduction. The previous works shed light on the few attempts that
target applying low-complexity NN models for SI cancellation in FD
systems. However, further enhancements in the complexity are required
to build energy-efficient NN-based cancelers, which can be suitable
for hardware implementation in mobile communication platforms. As
such, this study fills in this gap by providing efficient NN-based
SI cancelers, which achieve a similar cancellation performance to
that of the polynomial and the state-of-the-art NN-based cancelers
while attaining a remarkable complexity reduction.

Based on the aforementioned, in this article, two novel low-complexity
NN architectures referred to as the hybrid-convolutional recurrent
NN (HCRNN) and hybrid-convolutional recurrent dense NN (HCRDNN) are
proposed. The proposed NNs exploit hybrid hidden layers (e.g., convolutional,
recurrent, and/or dense) to efficiently model the memory effect and
non-linearity incorporated into the SI signal, with low complexity.
The key idea behind using hybrid layers is to build an NN model, which
makes use of the characteristics of the different layers employed
in its architecture. In particular, the proposed NNs exploit, in a
novel manner, the feature extraction characteristics of the convolutional
layer along with the sequence modeling capabilities of the recurrent
layer and/or the learning abilities of the dense layer in order to
model the SI with lower computational complexity than the polynomial
and the state-of-the-art NN-based cancelers. To the best of the authors\textquoteright{}
knowledge, applying hybrid-layers NN architectures for SI cancellation
has not been previously reported in the literature, and it is introduced
for the first time in this paper. More specifically, in the proposed
HCRNN, the input data containing the I/Q components of the input samples
is formulated into a two-dimensional (2D) graph for the sake of suitable
processing by the convolutional layer. The convolutional layer is
then applied to the 2D graph to extract the input features (e.g.,
memory effect and non-linearity) at a reduced network scale. Moreover,
a recurrent layer is then utilized to help in learning the temporal
behavior of the input signal from the output feature map of the convolutional
layer. In the proposed HCRDNN, a dense layer is added after the convolutional
and recurrent layers to build a deeper NN model with low computational
complexity. Working with hybrid-layers NN architectures enables adjusting
the hidden layers\textquoteright{} settings to achieve a certain cancellation
performance with a considerable computational complexity reduction. 

The contributions of this article are summarized as follows:
\begin{itemize}
\item Two novel hybrid-layers NN architectures, termed as the HCRNN and
HCRDNN, are proposed for the first time to model the SI in FD systems
with low computational complexity. In contrast to the state-of-the-art
NNs that directly apply the traditional dense or recurrent layers
for SI modeling, the proposed NNs exploit, in a novel manner, a combination
of hidden layers (e.g., convolutional, recurrent, and/or dense) in
order to achieve high learning capability while maintaining low computational
complexity. 
\item The computational complexity and memory requirements of the proposed
HCRNN and HCRDNN-based cancelers are derived in terms of the number
of floating-point operations (FLOPs) and network parameters, respectively,
and analyzed compared to those of the polynomial and the state-of-the-art
NN-based cancelers. 
\item The optimum settings for training the proposed HCRNN and HCRDNN architectures
(e.g., number of convolutional filters, filter size, number of neurons
in recurrent and dense layers, activation functions, learning rate,
batch size, and optimizer) are selected to achieve an acceptable cancellation
performance with a considerable computational complexity reduction.
\item Performance analysis of the two proposed NNs is provided in terms
of their prediction capabilities, mean square error (MSE), achieved
SI cancellation, computational complexity, and memory requirements.
Both NNs demonstrate excellent prediction capabilities in modeling
the interference in FD systems with reduced complexity. 
\end{itemize}
The rest of this article is organized as follows. Section \ref{sec:System-Model}
presents the FD transceiver system model. Section \ref{sec:Proposed-Hybrid-Layers-Neural}
introduces the proposed HCRNN and HCRDNN-based cancelers. In Section
\ref{sec:Computational-Complexity-Analysis}, the complexity of the
proposed NN architectures is analyzed, whereas in Section \ref{sec:Training-of-the Proposed HCRNN and HCRDNN},
the optimum settings for their training are selected. Finally, simulation
results, future research directions, and conclusions are presented
in Sections \ref{sec:Results-and-Discussion}, \ref{sec:Future-Research-Directions},
and \ref{sec:Conclusion}, respectively.\vspace{-2mm}

\section{\label{sec:System-Model} System Model}

\begin{figure*}[t]
\begin{centering}
\vspace{-8mm}
\subfloat[Detailed system model.]{\begin{raggedright}
\includegraphics[scale=0.24]{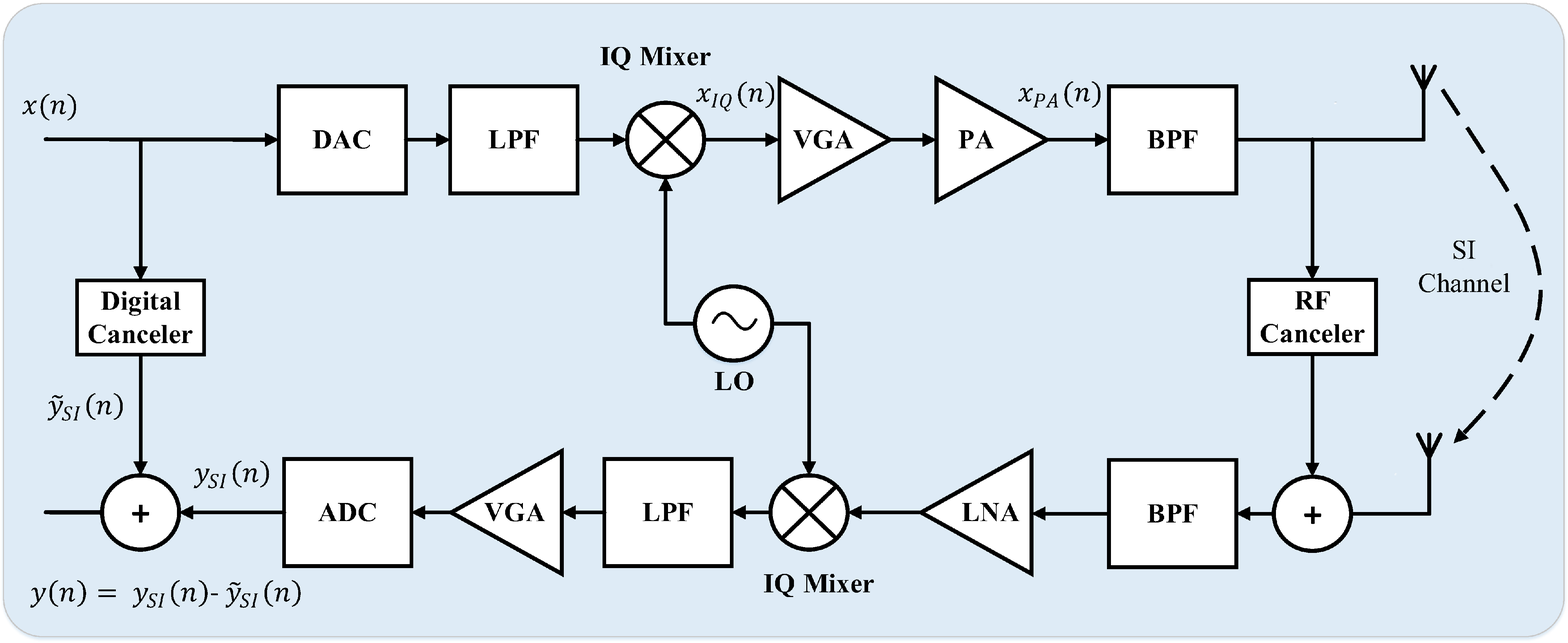}
\par\end{raggedright}
\centering{}} \subfloat[Digital canceler.]{\begin{raggedright}
\includegraphics[scale=0.24]{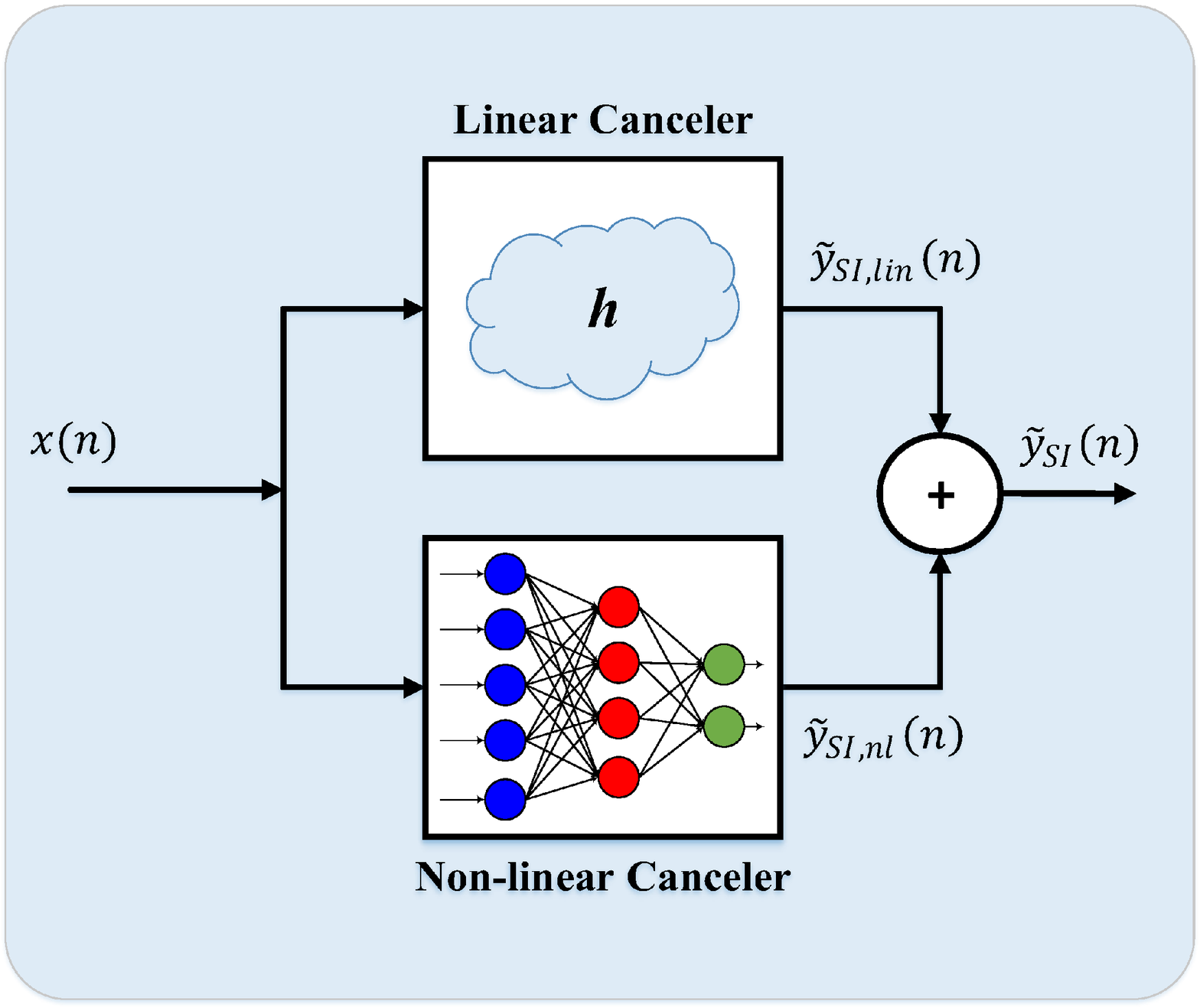}
\par\end{raggedright}
\raggedleft{}}
\par\end{centering}
\caption{\label{fig:Full-duplex-transceiver-system1}Full-duplex transceiver
system model.}
\end{figure*}

An FD transceiver consisting of a local transmitter, local receiver,
and two SI cancellation techniques is illustrated in Figs. \ref{fig:Full-duplex-transceiver-system1}(a)
and (b). Specifically, the FD system\textquoteright s design, shown
in Fig. \ref{fig:Full-duplex-transceiver-system1}(a), employs an
analog RF cancellation and a training-based digital cancellation in
order to suppress the SI signal to the receiver noise level. The RF
cancellation is applied at the first stage of the receiver chain to
prevent the SI signal from saturating the receiver's analog components
(e.g., LNA, VGA, and ADC). However, the digital cancellation is employed
after the ADC to remove the residual SI signal. 

Let us denote the digital transmitted samples before the DAC by $x(n)$,
with $n$ representing each sample index. The transmitted samples
are converted to analog, filtered, and up-converted to the carrier
frequency using the DAC, low pass filter (LPF), and IQ mixer, respectively.
The IQ mixer, undesirably, distorts the transmitted signal due to
the amplitude and gain imbalances between the I/Q components (i.e.,
IQ imbalance). Subsequently, the digital equivalent of the mixer's
output signal can be expressed as {[}\ref{D.-Korpi,-L.}{]}, {[}\ref{A.-Balatsoukas-Stimming,-"Non-li}{]}\vspace{-2mm}

\begin{equation}
x_{IQ}(n)=\frac{1}{2}(1+\psi e^{j\theta})\,x(n)+\frac{1}{2}(1-\psi e^{j\theta})\,x^{*}(n),\label{eq:1}
\end{equation}
where $\psi$ and $\theta$ represent the gain and phase imbalance
coefficients of the transmitter, respectively. The mixer's output
signal is then amplified by the PA, which further distorts the transmitted
signal due to its non-idealities. The PA's output signal can be expressed
using the conventional parallel-Hammerstein (PH) model, described
by (\ref{eq:52}) in the Appendix, as {[}\ref{D.-Korpi,-L.}{]}, {[}\ref{A.-Balatsoukas-Stimming,-"Non-li}{]},
{[}\ref{Volterra 1}{]}\vspace{-2mm}

\begin{equation}
x_{P\hspace{-0.5mm}A}(n)=\sum_{\underset{p\,odd}{p=1,}}^{P}\sum_{m=0}^{M_{P\hspace{-0.5mm}A}}h{}_{m,p}\thinspace x_{IQ}(n-m)^{\frac{p+1}{2}}x_{IQ}^{*}(n-m)^{\frac{p-1}{2}},\label{eq:2}
\end{equation}
where $h_{m,p}$ indicates the PA's impulse response. In addition,
$P$ and $M_{P\hspace{-0.5mm}A}$ represent the non-linearity order
and the PA's memory depth, respectively. 

The PA's output signal leaks to the receiver through the SI channel,
forming the SI signal. Accordingly, at the receiver side of the FD
node, there are three signals: an SI signal, a noise signal, and a
far-end desired signal from another FD node. In this work, we assume,
for simplicity of analysis, that there is no thermal noise, and there
are no far-end desired signals from any other FD nodes {[}\ref{A.-Balatsoukas-Stimming,-"Non-li}{]},
{[}\ref{Hardware implementation of neural self-interference cancellation}{]}.
As such, the residual SI signal after the RF cancellation process
is filtered, amplified, down-converted, and digitized using the band-pass
(BPF), LNA, IQ mixer, and ADC, respectively, and can be expressed
as {[}\ref{A.-Balatsoukas-Stimming,-"Non-li}{]}\vspace{-2mm}

\begin{equation}
y_{_{SI}}(n)=\sum_{\underset{p\,odd}{p=1,}}^{P}\sum_{q=0}^{p}\sum_{m=0}^{M-1}h{}_{m,q,p}\,x(n-m)^{q}x^{*}(n-m)^{p-q},\label{eq:3}
\end{equation}
where $h_{m,q,p}$ represents the impulse response of a channel, including
the composite effect of the PA, IQ mixer, and SI channel, whereas
$M$ indicates the memory effect incorporated into the input signal
by the PA and SI coupling channel. 

The estimation of the aforementioned channel is performed using the
least-squares (LS) approach as follows. Firstly, we formulate the
matrix $\boldsymbol{X}$ with dimensions $N\times M$ from the input
data $x$, where $N$ is the length of the training data. Then, the
output signal $\boldsymbol{y}$ with dimension $N\times1$ can be
obtained as $\boldsymbol{y=Xh}$, where $\boldsymbol{h}$ is a vector
of size $M\times1$. In the LS method, the aim is to minimize the
cost function $J(\boldsymbol{\boldsymbol{\hat{h}}})$, which can be
expressed as \vspace{-2mm}

\[
\hspace{-1.8cm}J(\boldsymbol{\boldsymbol{\hat{h}}})=\left\Vert \boldsymbol{y}-\boldsymbol{X}\boldsymbol{\hat{h}}\right\Vert ^{2}\hspace{-0.05cm}=(\boldsymbol{y}-\boldsymbol{X}\boldsymbol{\hat{h}})^{H}(\boldsymbol{y}-\boldsymbol{X}\boldsymbol{\hat{h}})
\]
\begin{equation}
\hspace{10.2mm}=\boldsymbol{y^{H}\boldsymbol{y}}-\boldsymbol{y^{H}\boldsymbol{X}\boldsymbol{\hat{h}}}-\boldsymbol{\hat{h}}^{H}\boldsymbol{\boldsymbol{X}^{H}}\boldsymbol{\boldsymbol{y}}+\boldsymbol{\hat{h}}^{H}\boldsymbol{\boldsymbol{X}^{H}}\boldsymbol{\boldsymbol{X}}\boldsymbol{\boldsymbol{\hat{h}}},\label{eq:4}
\end{equation}
where $\left(.\right)^{H}$denotes the Hermitian transpose operator.
By setting the derivative of this cost function with respect to $\boldsymbol{\hat{h}}$
to zero, the LS solution can be given as \vspace{-2mm}

\begin{equation}
\boldsymbol{\boldsymbol{\hat{h}}}=\left(\boldsymbol{X^{H}}\boldsymbol{X}\right)^{-1}\boldsymbol{X^{H}}\boldsymbol{y}.\label{eq:5}
\end{equation}

In the digital canceler, the goal is to estimate the distortion caused
by the imperfection of the transceiver hardware components and SI
channel to generate an accurate replica of the SI signal $\tilde{y}_{_{SI}}(n)$
at the receiver. This is attained by feeding the baseband transmitted
samples before the digital-to-analog conversion to a trainable-based
digital canceler in order to produce such a replica. This replica
is then subtracted from the SI signal after the ADC to remove the
interference, and the residual SI after the digital cancellation is
given by $y(n)=y_{_{SI}}(n)-\tilde{y}_{_{SI}}(n)$. The achieved SI
cancellation in the digital domain can be quantified in dB as \vspace{-2mm}

\begin{equation}
\mathbb{\mathcal{C}}_{dB}=10\log_{10}\left(\frac{\sum_{n}\left|y_{_{SI}}(n)\right|^{2}}{\sum_{n}\left|y(n)\right|^{2}}\right)\negmedspace.\label{eq:6}
\end{equation}
In this work, the digital canceler is formed by linear and non-linear
trainable-based cancelers, as depicted in Fig. \ref{fig:Full-duplex-transceiver-system1}(b).
The former is utilized to estimate the linear part of the SI signal
based on the conventional LS channel estimation {[}\ref{A.-Balatsoukas-Stimming,-"Non-li}{]},
whereas the latter is used to mimic the non-linear part of the SI
signal using an NN model. The SI signal is then reconstructed by combining
the linear and non-linear components as follows: \vspace{-2mm}

\begin{equation}
\tilde{y}_{_{SI}}(n)=\tilde{y}_{_{SI,lin}}(n)+\tilde{y}_{_{SI,nl}}(n),\label{eq:7}
\end{equation}
where $\tilde{y}_{_{SI,lin}}(n)$ is the linear part of the SI signal,
which can be obtained by substituting $p=1$ and $q=1$ in (\ref{eq:3})
as 

\vspace{-2mm}

\begin{equation}
\tilde{y}_{_{SI,lin}}(n)=\sum_{m=0}^{M-1}h{}_{m,1,1}\,x(n-m),\label{eq:8}
\end{equation}
while $\tilde{y}_{_{SI,nl}}(n)$ is the non-linear part, which can
be given as

\vspace{-2mm}

\begin{equation}
\tilde{y}_{_{SI,nl}}(n)=g\left\{ x\left(n\right),x\left(n-1\right),...,x\left(n-M+1\right)\right\} ,\label{eq:9}
\end{equation}
where $g\left\{ .\right\} $ represents the NN mapping function.

\section{\label{sec:Proposed-Hybrid-Layers-Neural}Proposed Hybrid-Layers
NN Architectures }

\begin{figure*}
\begin{centering}
\vspace{-8mm}
\includegraphics[scale=0.17]{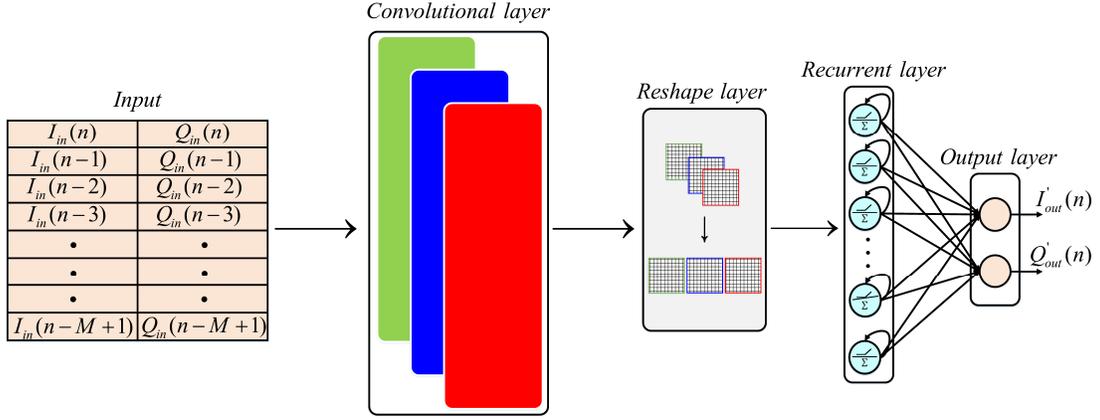}
\par\end{centering}
\centering{}\caption{\label{fig:Proposed-Hybrid-Conventional-Rec}Proposed HCRNN architecture.}
\end{figure*}

The main function of the NN-based canceler is to provide an accurate
behavior model to mimic the non-linearities and memory effect attributed
to the input signal. As such, to account for such effects, TDNN or
RNN-based models were employed in the literature. While the aforementioned
models attain good modeling capabilities for approximating the SI
in FD systems, their complexity is a considerable issue that should
be addressed. A good choice for designing a low-complexity NN model
is to exploit the high feature extraction along with the parameter
reduction capabilities of the convolutional layer. Further, adding
a recurrent layer with a sufficient number of neurons after the convolutional
layer can enhance the modeling capabilities with no much increase
in the computational complexity. Lastly, exploiting an additional
dense layer after the recurrent layer can help to add another degree
of freedom for adapting the NN settings in order to achieve the best
compromise between the model performance and the computational complexity.
In the next subsections, we will show how the proposed NN architectures
exploit the aforementioned layers in order to build low-complexity
NN-based cancelers.\vspace{-2mm}

\subsection{HCRNN Architecture }

\begin{figure*}
\begin{centering}
\vspace{-5mm}
\includegraphics[scale=0.135]{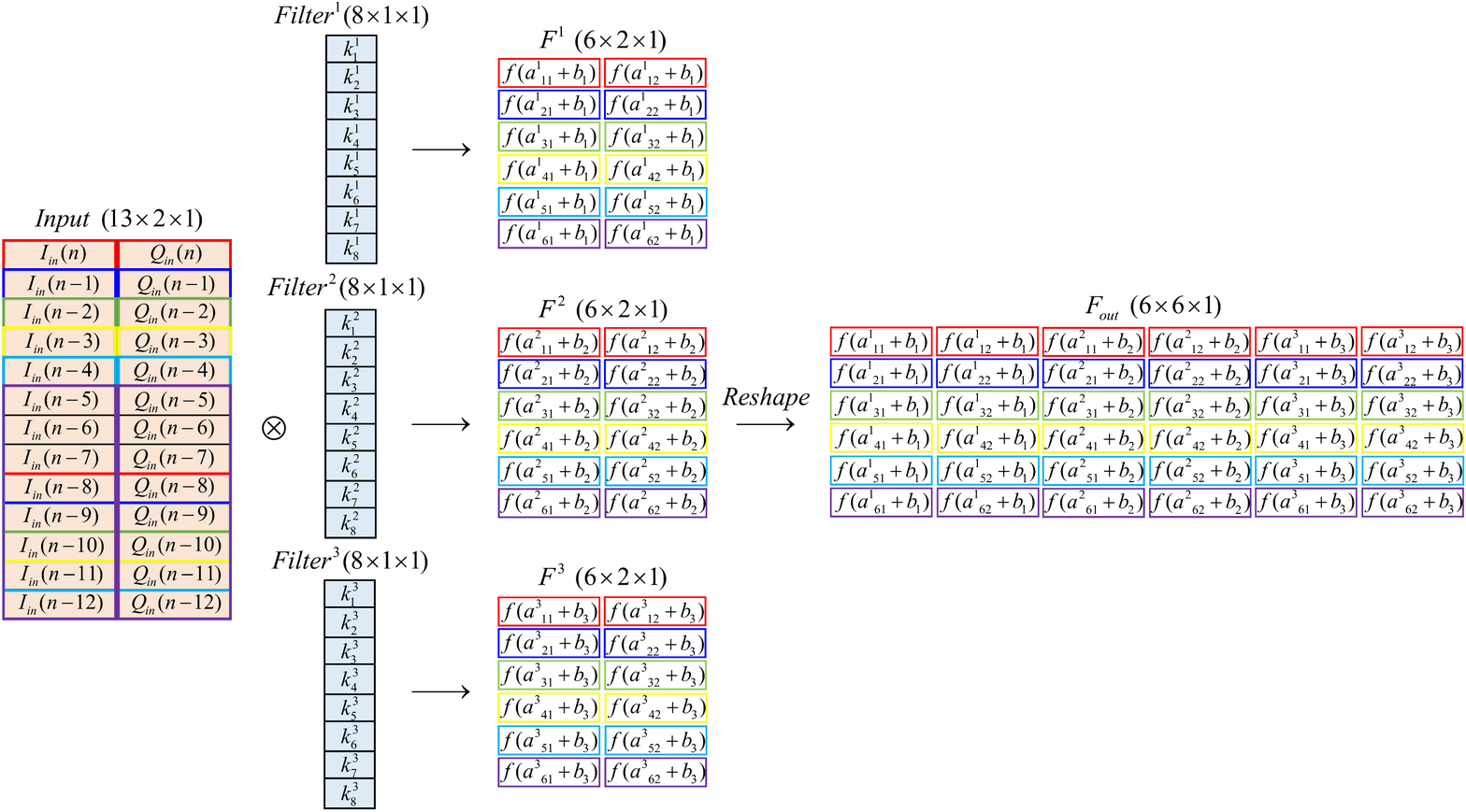}
\par\end{centering}
\centering{}\caption{\label{fig:Convolutional-layer-opertion}Example of the basic operation
of the proposed HCRNN.}
\end{figure*}

The proposed HCRNN architecture is shown in Fig. \ref{fig:Proposed-Hybrid-Conventional-Rec}.
In the HCRNN, hybrid hidden layers are employed to detect the memory
effect and non-linearity incorporated into the input signal due to
the impairments of the FD transceiver and SI channel. The HCRNN is
arranged in five layers: the input layer, convolutional layer, reshape
layer, recurrent layer, and output layer. At the input layer, the
input data is formulated into a 2D graph consisting of the I/Q components
of the instantaneous and delayed versions of the input samples. Arranging
the data into a 2D graph allows the input samples to be in an appropriate
form that can be efficiently processed by the next convolutional layer.
The convolutional layer, the first hidden layer of the HCRNN, is applied
to the 2D grid in order to detect the input signal's features with
a reduced network scale. Applying the convolutional layer to the input
graph comes with a considerable reduction in the computational complexity
due to the weight-sharing characteristics and dimensionality reduction
capabilities of the convolutional filters {[}\ref{Paper of (11) }{]}.
The convolutional layer's output is then reordered using a reshape
layer, to be processed by the next recurrent layer, which constitutes
the second hidden layer of the HCRNN. The aim of the recurrent layer
is to assist in learning the temporal behavior of the input signal
from the localized feature map of the convolutional layer. Finally,
at the output layer of the HCRNN, the I/Q components of the SI signal
are estimated. 

An example showing the basic operation of the proposed HCRNN, using
$M=13$, three convolutional filters, and $8\times1\times1$ filter
size, is illustrated in Fig. \ref{fig:Convolutional-layer-opertion}.
At the convolutional layer, the input data is convolved with the filters,
acting as optimizable-feature extractors, in order to detect the patterns
of the input attributes. The output feature map of the $l^{th}$ filter
after applying the convolution operation can be expressed as

\begin{equation}
\boldsymbol{F}_{c}^{l}=\begin{bmatrix}I_{in}(n) & Q_{in}(n)\\
I_{in}(n-1) & Q_{in}(n-1)\\
I_{in}(n-2) & Q_{in}(n-2)\\
. & .\\
. & .\\
I_{in}(n-M+1) & Q_{in}(n-M+1)
\end{bmatrix}\otimes\begin{bmatrix}k_{1,1}^{l} & k_{1,S}^{l}\\
k_{2,1}^{l} & k_{2,S}^{l}\\
k_{3,1}^{l} & k_{3,S}^{l}\\
. & .\\
. & .\\
k_{R,1}^{l} & k_{R,S}^{l}
\end{bmatrix},\label{eq:10}
\end{equation}
where $l=1,2,...,L$, with $L$ denoting the number of convolutional
filters. $I_{in}(n)$ and $Q_{in}(n)$ represent the I/Q components
of the current sample $x(n)$, respectively, while $[I_{in}(n-1),...,I_{in}(n-M+1)]$
and $[Q_{in}(n-1),...,Q_{in}(n-M+1)]$ indicate the I/Q components
of the delayed samples, respectively. In addition, $R\times S\times Z$
and $k_{i,j}^{l}$ represent the dimensions and $(i^{th},j^{th})$
entry of the $l^{th}$ convolutional filter, respectively. Finally,
$\otimes$ denotes the convolution operation. It is worth mentioning
that for all convolutional filters, $R\in\left\{ 1,2,...,M\right\} $,
$S\in\left\{ 1,2\right\} $, and $Z=1$ since the input data is arranged
into a 2D graph consisting of two columns to represent the I/Q components
of the instantaneous and delayed versions of the input samples.

After performing the convolution, a bias term is added, and a non-linear
activation function is applied to each element of $\boldsymbol{F}_{c}^{l}$.
Accordingly, the output feature map of the $l^{th}$ convolutional
filter can be expressed as \vspace{-2mm}

\begin{equation}
\boldsymbol{F}^{l}=\begin{bmatrix}f^{c}(a_{1,1}^{l}+b_{l}) & f^{c}(a_{1,C}^{l}+b_{l})\\
f^{c}(a_{2,1}^{l}+b_{l}) & f^{c}(a_{2,C}^{l}+b_{l})\\
f^{c}(a_{3,1}^{l}+b_{l}) & f^{c}(a_{3,C}^{l}+b_{l})\\
. & .\\
. & .\\
f^{c}(a_{B,1}^{l}+b_{l}) & f^{c}(a_{B,C}^{l}+b_{l})
\end{bmatrix},\label{eq:11}
\end{equation}
where $f^{c}(.)$ denotes the convolutional layer's activation function,
while $b_{l}$ represents the bias term associated with the $l^{th}$
convolutional filter. $B\times C$ indicates the output feature map's
dimensions, whereas $a_{i,j}^{l}$ denotes the $(i^{th},j^{th})$
entry of the feature map just after applying the convolution. With
unity-stride and without zero-padding, $a_{i,j}^{l}$ can be given
as \vspace{-2mm}

\begin{equation}
a_{i,j}^{l}=\sum_{r=0}^{R-1}\sum_{s=0}^{S-1}\boldsymbol{X}_{r,s+j}\boldsymbol{K}_{r,s}^{l},\label{eq:12}
\end{equation}
where $\boldsymbol{X}$ and $\boldsymbol{K}^{l}$ represent the input
and the $l^{th}$ convolutional filter matrices with entries described
in (\ref{eq:10}), respectively.

For efficient processing of the forwarded data through the network,
the output feature maps of all filters are reshaped before they are
passed to the recurrent layer. More specifically, in the output feature
map of each filter, there are dependencies between each column's elements
due to the temporal behavior existing in the input data. Therefore,
in the proposed HCRNN, the output feature maps of all filters are
reformulated using the reshape layer in order to take these dependencies
into account and pass them in a sequence to the recurrent layer. Reshaping
the feature maps with this mechanism enables the recurrent layer to
detect the aforementioned sequence for proper modeling of the system's
temporal behavior. Based on this, the resultant feature map after
the reshaping process can be expressed as \vspace{-2mm}

\begin{equation}
\boldsymbol{F}_{out}=\begin{bmatrix}\boldsymbol{F}^{1} & \boldsymbol{F}^{2} & ... & \boldsymbol{F}^{L}\end{bmatrix},\label{eq:13}
\end{equation}
where $\boldsymbol{F}^{1}$, $\boldsymbol{F}^{2}$, and $\boldsymbol{F}^{L}$
represent the output feature maps of the $1^{st}$, $2^{nd}$, and
$L^{th}$ convolutional filters, respectively. The reshaped feature
map is then passed to the recurrent layer, and the output at any time
step $t$ can be expressed as \vspace{-2mm}

\begin{equation}
\boldsymbol{y}_{r}(t)=f^{r}\left(\boldsymbol{f}_{out}(t)\boldsymbol{W}_{x}+\boldsymbol{y}_{r}(t-1)\boldsymbol{W}_{y}+\boldsymbol{b}_{r}\right),\label{eq:14}
\end{equation}
where $\boldsymbol{y}_{r}(t)\in\mathbb{R}^{1\times n_{hr}}$ represents
the recurrent layer output at any time step $t$, with $n_{hr}$ denoting
the number of recurrent layer's neurons. Similarly, $\boldsymbol{y}_{r}(t-1)\in\mathbb{R}^{1\times n_{hr}}$
indicates the recurrent layer's output at the previous time step $t-1$.
$\boldsymbol{f}_{out}(t)\in\mathbb{R}^{1\times n_{i}}$ represents
a row vector of $\boldsymbol{F}_{out}$, which is passed to the recurrent
layer at time step $t$, with $n_{i}$ as the number of input features.
$\boldsymbol{W}_{x}\in\mathbb{R}^{n_{i}\times n_{hr}}$ denotes the
weight matrix for the connections between the input and hidden units
at the current time step. Further, $\boldsymbol{W}_{y}\in\mathbb{R}^{n_{hr}\times n_{hr}}$
indicates the weight matrix for the feedback connections from the
hidden units at the previous time step. Finally, $f^{r}(.)$ represents
the activation function operation of the recurrent layer, whereas
$\boldsymbol{b}_{r}\in\mathbb{R}^{1\times n_{hr}}$ indicates a row
vector containing the bias terms of the recurrent layer neurons. 

The recurrent layer's output is then passed to the output layer, which
is formed by a fully-connected (dense) layer containing two neurons.
The output layer neurons are utilized to map the extracted features
by the recurrent layer to the final output (i.e., estimated I/Q components
of the SI signal) as follows:\vspace{-2mm}

\begin{equation}
I_{out}^{'}(n)=f^{o}\left(\sum_{i=1}^{n_{hr}}\boldsymbol{W}_{o}^{1,i}\mathbf{\boldsymbol{y}}_{r}^{i}+\boldsymbol{b}_{o}^{1}\right),\label{eq:15}
\end{equation}
\vspace{-2mm}

\begin{equation}
Q_{out}^{'}(n)=f^{o}\left(\sum_{i=1}^{n_{hr}}\boldsymbol{W}_{o}^{2,i}\boldsymbol{y}_{r}^{i}+\boldsymbol{b}_{o}^{2}\right),\label{eq:16}
\end{equation}
where $f^{o}(.)$ represents the output layer's activation function,
while $\{\boldsymbol{W}_{o}^{1,i},\boldsymbol{b}_{o,}^{1}\}$ and
$\{\boldsymbol{W}_{o}^{2,i},\boldsymbol{b}_{o}^{2}\}$ indicate the
weight and bias terms associated with the first and second neurons
of the output layer, respectively.

\subsection{HCRDNN Architecture }

\begin{figure*}
\begin{centering}
\vspace{-2mm}
\includegraphics[scale=0.17]{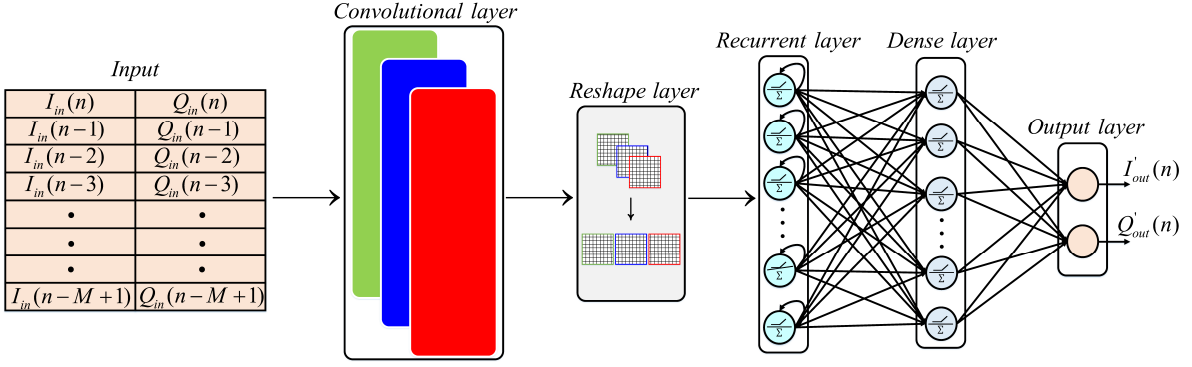}
\par\end{centering}
\centering{}\caption{\label{fig:(HCRDNN)}Proposed HCRDNN architecture.}
\end{figure*}

The network architecture of the proposed HCRDNN is depicted in Fig.
\ref{fig:(HCRDNN)}. The key difference between the HCRDNN and HCRNN
architectures is that an additional dense layer is employed after
the convolutional and recurrent layers as increasing the number of
hidden layers can enable building more complex NN models with reduced
complexity {[}\ref{Paper of (11) }{]}. Additionally, adding a dense
layer to the HCRDNN architecture provides another degree of freedom
to adapt the NN settings in order to achieve the optimum cancellation-complexity
trade-off. In particular, in the HCRDNN, the convolutional layer's
hyper-parameters (e.g., number of filters, filter size) and the number
of neurons in the recurrent and dense layers are jointly adjusted
to achieve a considerable cancellation performance with a significant
computational complexity reduction. 

The output of the $u^{th}$ neuron at the dense layer of the HCRDNN
can be expressed as \vspace{-2mm}

\begin{equation}
y_{d}^{u}=f^{d}\left(\sum_{i=1}^{n_{hr}}\boldsymbol{W}_{d}^{u,i}\boldsymbol{y}_{r}^{i}+\boldsymbol{b}_{d}^{u}\right),\label{eq:17}
\end{equation}
where $f^{d}(.)$ indicates the activation function of the dense layer,
while $\{\boldsymbol{W}_{d}^{u,i},\boldsymbol{b}_{d,}^{u}\}$ represent
the weight and bias terms associated with the $u^{th}$ neuron in
the dense layer, respectively. The dense layer's output is then passed
to the output layer to estimate the I/Q components of the SI signal
as (\ref{eq:15}) and (\ref{eq:16}).

After predicting $I_{out}^{'}(n)$ and $Q_{out}^{'}(n)$, the MSE
is calculated as a cost function to measure how well the proposed
NN models predict the actual outputs as follows: \vspace{-2mm}

\begin{equation}
E=\hspace{-0.5mm}\frac{1}{2N}\hspace{-0.5mm}\sum_{n=1}^{N}\hspace{-1mm}\left(\hspace{-0.5mm}\left[I_{out}(n)\hspace{-0.5mm}-\hspace{-0.5mm}I_{out}^{'}(n)\right]^{2}\hspace{-1mm}+\hspace{-0.5mm}\left[Q_{out}(n)\hspace{-0.5mm}-\hspace{-0.5mm}Q_{out}^{'}(n)\right]^{2}\right),\label{eq:18}
\end{equation}
where $\{I_{out}(n),$ $I_{out}(n)\}$ and $\{I_{out}^{'}(n),$ $Q_{out}^{'}(n)\}$
represent the actual and predicted values of the I/Q components of
the SI signal, receptively, whereas $N$ denotes the number of training
observations, as stated before in Section \ref{sec:System-Model}.
During the HCRNN and HCRDNN models' training, the convolutional, recurrent,
and dense layers' weights and biases are set to minimize this cost
function. 

After the training process, the proposed NNs are employed to provide
non-linear SI cancellation as part of the digital cancellation in
the FD transceiver, as illustrated in Fig. \ref{fig:Full-duplex-transceiver-system1}(b). 

\section{\label{sec:Computational-Complexity-Analysis} Complexity Analysis}

In this section, the computational complexity and memory requirements
of the polynomial and proposed NN-based cancelers are analyzed in
terms of the number of computations and memory usage of the linear
and non-linear cancelers required to provide the total SI cancellation
(i.e., the summation of linear and non-linear cancellations). In particular,
in this work, the computational complexity is assessed in terms of
the number of FLOPs used to perform the linear and non-linear cancellation
processes. Moreover, the memory usage requirements are assessed in
terms of the number of stored parameters utilized to achieve the total
cancellation. In this analysis, we focus on evaluating the complexity
of the polynomial and proposed NN-based cancelers in the real-time
inference stage since powerful processing units can be employed to
perform the training process offline.

\subsection{Linear Canceler Complexity }

In this subsection, we calculate the number of FLOPs required to implement
the linear canceler in terms of the number of RV multiplications and
additions. From (\ref{eq:8}), the linear canceler requires $M$ complex
multiplications and $M-1$ complex additions to perform the linear
cancellation. Using the assumption that each complex addition is implemented
by two real additions, and each complex multiplication is executed
using three real multiplications and five real additions {[}\ref{Hardware implementation of neural self-interference cancellation}{]},
the number FLOPs of the linear canceler can be expressed as \vspace{-2mm}

\begin{equation}
\mathscr{\mathcal{F}_{\mathit{lin}}}=\underset{\chi_{mul,lin}^{\Re}}{\underbrace{3M}}+\underset{\text{\ensuremath{\chi_{add,lin}^{\mathfrak{\mathbb{\Re}}}}}}{\underbrace{7M-2}},\label{eq:19}
\end{equation}
where $\chi_{mul,lin}^{\Re}$ and $\chi_{add,lin}^{\mathfrak{\mathbb{\Re}}}$
represent the number of real multiplications and additions required
for the linear canceler, respectively, while $M$ denotes the memory
effect incorporated into the input signal by the PA and SI channel,
as stated before. Similarly, from (\ref{eq:8}), the memory requirements
of the linear canceler is evaluated as a function of the number of
stored parameters, which can be expressed as 

\vspace{-6mm}

\begin{equation}
\mathcal{P}_{lin}=2M.\label{eq:20}
\end{equation}

\subsection{Non-Linear Canceler Complexity }

In this subsection, the number of FLOPs and parameters required for
non-linear cancellation using polynomial and NN-based cancelers are
analyzed. 

\subsubsection{Polynomial-based Canceler Complexity}

Using the previous assumptions, it can be deduced from (\ref{eq:3})
that the number of RV multiplications and additions required for the
non-linear polynomial-based canceler can be expressed as {[}\ref{A.-Balatsoukas-Stimming,-"Non-li}{]},
{[}\ref{Hardware implementation of neural self-interference cancellation}{]}
\vspace{-2mm}

\begin{equation}
\chi_{mul,poly}^{\mathfrak{\mathbb{\Re}}}=3M\left\{ \left(\frac{P+1}{2}\right)\left(\frac{P+1}{2}+1\right)-1\right\} ,\label{eq:21}
\end{equation}
\vspace{-2mm}

\begin{equation}
\chi_{add,poly}^{\mathfrak{\mathbb{\Re}}}=7M\left\{ \left(\frac{P+1}{2}\right)\left(\frac{P+1}{2}+1\right)-1\right\} .\label{eq:22}
\end{equation}
Similarly, the number of parameters of the non-linear polynomial-based
canceler can be expressed as {[}\ref{A.-Balatsoukas-Stimming,-"Non-li}{]},
{[}\ref{Hardware implementation of neural self-interference cancellation}{]}
\vspace{-2mm}

\begin{equation}
\thinspace\thinspace\thinspace\thinspace\thinspace\thinspace\thinspace\thinspace\thinspace\thinspace\thinspace\thinspace\thinspace\thinspace\thinspace\mathcal{P}_{poly}=2M\left\{ \left(\frac{P+1}{2}\right)\left(\frac{P+1}{2}+1\right)-1\right\} .\label{eq:23}
\end{equation}

\subsubsection{NN-based Canceler Complexity}

For the proposed non-linear HCRNN-based canceler, the number of FLOPs
can be expressed as\vspace{-2mm}

\begin{equation}
\mathscr{\mathcal{F}_{\mathit{HCRNN}}}=\mathcal{F^{\mathit{c}}}+\mathcal{F^{\mathit{r}}}+\mathcal{F^{\mathit{o}}},\label{eq:24}
\end{equation}
where $\mathcal{F}^{\mathit{c}}$, $\mathcal{F}^{\mathit{r}}$, and
$\mathcal{F}^{\mathit{o}}$ represent the number of FLOPs required
for the convolutional, recurrent, and output layers of the HCRNN,
respectively. Firstly, for the convolutional layer, $\mathcal{F}^{\mathit{c}}$
can be expressed as \vspace{-2mm}

$\vphantom{}$

$\underset{\text{ \thinspace\thinspace\thinspace\thinspace\thinspace\thinspace\thinspace\thinspace\thinspace\thinspace\thinspace\thinspace\thinspace\thinspace\thinspace\thinspace\thinspace\thinspace\thinspace\thinspace\thinspace\thinspace\thinspace}\textrm{Ter\ensuremath{\textrm{m}_{1}^{c}}}}{\mathcal{F^{\mathit{c}}}=\underbrace{\left(2\left\{ R\times S\times Z\right\} -1\right)\left(B\times C\times L\right)}}$

\begin{equation}
\thinspace\thinspace\thinspace\thinspace\thinspace\thinspace\thinspace\thinspace\thinspace\thinspace\thinspace\thinspace\thinspace\thinspace\thinspace\thinspace\thinspace\thinspace\thinspace\thinspace\thinspace\thinspace\thinspace\thinspace\thinspace\thinspace\thinspace\thinspace\thinspace\thinspace\thinspace\thinspace\thinspace\thinspace\thinspace\thinspace\thinspace\underset{\text{\thinspace\thinspace\thinspace\thinspace\thinspace\thinspace\thinspace\thinspace\thinspace\thinspace\thinspace}\textrm{Ter\ensuremath{\textrm{m}_{2}^{c}}}}{+\underbrace{\mathcal{F^{\mathit{act}}}\times\left(B\times C\times L\right)}}+\underset{\text{\thinspace\thinspace}\textrm{Ter\ensuremath{\textrm{m}_{3}^{c}}}}{\underbrace{\left(B\times C\times L\right)}},\label{eq:25}
\end{equation}
where $\textrm{Ter}\textrm{m}_{1}^{c}$ represents the number of RV
multiplications and additions required to convolve the 2D graph of
the input data with $L$ convolutional filters. $\textrm{Ter}\textrm{m}_{2}^{c}$
denotes the number of RV operations (multiplications + additions)
required for applying the activation functions at each element of
the output feature map after the convolution process. $\textrm{Ter}\textrm{m}_{3}^{c}$
indicates the number of real additions required for adding the bias
values. Finally, $\mathcal{F^{\mathit{act}}}$ represents the number
of real operations needed to evaluate each activation function, which
mainly depends on the activation function type.

In this work, we consider the rectified linear unit ($ReLU$), $Sigmoid$,
and hyperbolic tangent ($tanh$) activation functions, which are defined
as {[}\ref{Paper of (11) }{]}
\begin{subequations}
\label{A}
\begin{align}
ReLU(z) & =\max(0,z),\label{eq: A_1}\\
Sigmoid(z) & =\frac{1}{e^{-z}+1},\label{eq: A_2}\\
tanh(z) & =\frac{e^{z}-e^{-z}}{e^{z}+e^{-z}}.\label{eq: A_3}
\end{align}
 
\end{subequations}

\hspace*{-0.33cm}By assuming that each real operation (e.g., multiplication,
division, addition, subtraction, and exponentiation) costs one FLOP
{[}\ref{Levenberg-marquardt learning neural network for adaptive predistortion }{]},
{[}\ref{A hybrid forward algorithm for RBF neural network construction}{]},
the number of FLOPs required to implement each of the aforementioned
activation functions can be expressed as\vspace{-2mm}

\begin{equation}
\mathcal{F^{\mathit{act}}}=\begin{cases}
1, & \textrm{ \thinspace\thinspace\thinspace\thinspace if }ReLU\\
4, & \textrm{ \thinspace\thinspace\thinspace\thinspace if }Sigmoid\\
6, & \textrm{ \thinspace\thinspace\thinspace\thinspace if }tanh
\end{cases}.\label{eq:29}
\end{equation}
Secondly, the number of FLOPs required for the recurrent layer can
be expressed as \vspace{-2mm}

\begin{equation}
\underset{\text{ }}{\underset{\text{ \thinspace\thinspace\thinspace\thinspace\thinspace\thinspace\thinspace\thinspace\thinspace\thinspace\thinspace\thinspace\thinspace\thinspace\thinspace\thinspace\thinspace\thinspace\thinspace\thinspace\thinspace\thinspace\thinspace\textrm{Ter\ensuremath{\textrm{m}_{1}^{r}}}}}{\mathcal{F^{\mathit{r}}}=\underbrace{2n_{hr}\left(n_{i}+n_{hr}-\frac{1}{2}\right)}}}\underset{\text{\thinspace\thinspace\thinspace\thinspace\thinspace\thinspace\thinspace\thinspace\thinspace\thinspace\thinspace\thinspace\thinspace\textrm{Ter\ensuremath{\textrm{m}_{2}^{r}}}}}{+\underbrace{\mathcal{F^{\mathit{act}}}\times n_{hr}}}+\underset{\text{\thinspace\thinspace\textrm{Ter\ensuremath{\textrm{m}_{3}^{r}}}}}{\underbrace{n_{hr}}},\label{eq:30}
\end{equation}
where $\textrm{Ter}\textrm{m}_{1}^{r}$ represents the number of RV
operations associated with multiplying the weight matrices $\boldsymbol{W}_{x}$
and $\boldsymbol{W}_{y}$ in (\ref{eq:14}) with their corresponding
inputs, while $\textrm{Ter}\textrm{m}_{2}^{r}$ and $\textrm{Ter}\textrm{m}_{3}^{r}$
indicate the number of real operations required for applying the activation
functions and adding the biases at each neuron of the recurrent layer,
respectively. 

Finally, the output layer's FLOPs can be expressed as \vspace{-2mm}

\begin{equation}
\underset{\text{\thinspace\thinspace\thinspace\thinspace\thinspace\thinspace\thinspace\thinspace\thinspace\thinspace\thinspace\thinspace\thinspace\thinspace\thinspace\thinspace\thinspace\thinspace\thinspace\thinspace\thinspace\thinspace\thinspace\thinspace\thinspace\thinspace\textrm{Ter\ensuremath{\textrm{m}_{1}^{o}}}}}{\mathcal{F^{\mathit{o}}}=\underbrace{n_{ho}\left(2n_{hr}-1\right)}}\underset{\text{\thinspace\thinspace\thinspace\thinspace\thinspace\thinspace\thinspace\thinspace\thinspace\thinspace\thinspace\textrm{Ter\ensuremath{\textrm{m}_{2}^{o}}}}}{+\underbrace{\mathcal{F^{\mathit{act}}}\times n_{ho}}}+\underset{\text{\thinspace\thinspace\textrm{Ter\ensuremath{\textrm{m}_{3}^{o}}}}}{\underbrace{n_{ho}}},\label{eq:31}
\end{equation}
where $n_{ho}$ represents the number of output layer's neurons. $\textrm{Ter}\textrm{m}_{1}^{o}$
denotes the number of RV operations required to calculate the weighted
sum of inputs at the output layer neurons. $\textrm{Ter}\textrm{m}_{2}^{o}$
and $\textrm{Ter}\textrm{m}_{3}^{o}$ indicate the number of real
operations associated with employing the activation functions and
adding the bias terms at each neuron of the output layer, respectively.

The number of parameters of the proposed HCRNN architecture can be
given as \vspace{-2mm}

\begin{equation}
\mathcal{P}_{HCRNN}=\mathcal{P^{\mathit{c}}}+\mathcal{P^{\mathit{r}}+\mathcal{P^{\mathit{o}}}},\label{eq:32}
\end{equation}
where $\mathcal{P^{\mathit{c}}}$, $\mathcal{P^{\mathit{r}}}$, and
$\mathcal{P^{\mathit{o}}}$ represent the number of parameters of
the convolutional, recurrent, and output layers, respectively, which
can be expressed as 

\begin{subequations}
\label{A-1}
\begin{align}
\mathcal{P^{\mathit{c}}} & =L\left(R\times S\times Z+1\right),\label{eq: A_1-1}\\
\mathcal{P}^{\mathit{r}} & =n_{hr}\left(n_{i}+n_{hr}+1\right),\label{eq: A_2-1}\\
\mathcal{P^{\mathit{o}}} & =n_{ho}\left(n_{hr}+1\right).\label{eq: A_3-1}
\end{align}
\end{subequations}

Using the same mathematical formulation, the number of FLOPs of the
proposed HCRDNN can be calculated as \vspace{-2mm}

\begin{equation}
\mathscr{\mathcal{F}_{\mathit{HCRDNN}}}=\mathcal{F^{\mathit{c}}}+\mathcal{F^{\mathit{r}}}+\mathcal{F^{\mathit{d}}}+\mathcal{F^{\mathit{o}}},\label{eq:36}
\end{equation}
where $\mathcal{F}^{\mathit{c}}$ and $\mathcal{F}^{\mathit{r}}$
represent the convolutional and recurrent layers' FLOPs, which can
be determined using (\ref{eq:25}) and (\ref{eq:30}), respectively.
$\mathcal{F}^{\mathit{o}}$ indicates the output layer's FLOPs, which
can be calculated by replacing $n_{hr}$ by $n_{hd}$ in (\ref{eq:31}),
whereas $\mathcal{F}^{\mathit{d}}$ denotes the dense layer's FLOPs,
which can be expressed as \vspace{-2mm}

\begin{equation}
\underset{\text{\thinspace\thinspace\thinspace\thinspace\thinspace\thinspace\thinspace\thinspace\thinspace\thinspace\thinspace\thinspace\thinspace\thinspace\thinspace\thinspace\thinspace\thinspace\thinspace\thinspace\thinspace\thinspace\thinspace\thinspace\thinspace\thinspace\textrm{Ter\ensuremath{\textrm{m}_{1}^{d}}}}}{\mathcal{F^{\mathit{d}}}=\underbrace{n_{hd}\left(2n_{hr}-1\right)}}\underset{\text{\thinspace\thinspace\thinspace\thinspace\thinspace\thinspace\thinspace\thinspace\thinspace\thinspace\thinspace\textrm{Ter\ensuremath{\textrm{m}_{2}^{d}}}}}{+\underbrace{\mathcal{F^{\mathit{act}}}\times n_{hd}}}+\underset{\text{\thinspace\thinspace\textrm{Ter\ensuremath{\textrm{m}_{3}^{d}}}}}{\underbrace{n_{hd}}},\label{eq:37}
\end{equation}
where $\textrm{Ter}\textrm{m}_{1}^{d}$, $\textrm{Ter}\textrm{m}_{2}^{d}$,
and $\textrm{Ter}\textrm{m}_{3}^{d}$ represent the number of real
operations required for calculating the weighted sum of inputs, applying
the activation functions, and summing the biases at the dense layer
neurons, respectively.

The number of parameters of the HCRDNN can be given as\vspace{-2mm}

\begin{equation}
\mathcal{P}_{HCRDNN}=\mathcal{P^{\mathit{c}}}+\mathcal{P^{\mathit{r}}+\mathcal{P^{\mathit{d}}}}+\mathcal{P^{\mathit{o}}},\label{eq:38}
\end{equation}
where $\mathcal{P}^{\mathit{c}}$ and $\mathcal{P}^{\mathit{r}}$
represent the convolutional and recurrent layers' parameters given
by (\ref{eq: A_1-1}) and (\ref{eq: A_2-1}), respectively, while
$\mathcal{P}^{\mathit{o}}$ indicates the output layer's parameters,
which can be evaluated by replacing $n_{hr}$ by $n_{hd}$ in (\ref{eq: A_3-1}).
Lastly, $\mathcal{P}^{\mathit{d}}=n_{hd}\left(n_{hr}+1\right)$ is
the number of dense layer parameters. 

For the RV-TDNN {[}\ref{A.-Balatsoukas-Stimming,-"Non-li}{]}, {[}\ref{A.-T.-Kristensen,}{]},
the number of FLOPs and parameters can be expressed as\vspace{-2mm}

\begin{equation}
\mathscr{\mathcal{F}_{\mathit{RV-TDNN}}}=\sum_{j=2}^{\mathcal{L}}2n_{j-1}n_{j}+\mathcal{F^{\mathit{act,j}}}n_{j},\label{eq:39}
\end{equation}
\vspace{-2mm}

\begin{equation}
\mathcal{P}_{RV-TDNN}=\sum_{j=2}^{\mathcal{L}}n_{j-1}n_{j}+n_{j},\label{eq:40}
\end{equation}
where $\mathcal{L}$ indicates the number of layers of the RV-TDNN,
including the input, hidden, and output layers. $n_{j}$ is the number
of neurons in the $j^{th}$ layer, with $n_{1}$ and $n_{\mathcal{L}}$
representing the number of input and output layers' neurons, respectively.
$\mathcal{\mathcal{F^{\mathit{act,j}}}}$ denotes the number of operations
required to apply the activation function at each of the $j^{th}$
layer's neurons.

Similarly, for the RNN {[}\ref{A.-T.-Kristensen,}{]}, the number
of FLOPs and parameters can be given by 

\vspace{-2mm}

\begin{equation}
\mathscr{\mathcal{F}_{\mathit{RNN}}}=\sum_{j=2}^{\mathcal{L}}2n_{j-1}n_{j}+2n_{j}^{2}+\mathcal{F^{\mathit{act,j}}}n_{j}-2n_{\mathcal{L}}^{2},\label{eq:41}
\end{equation}
\vspace{-2mm}

\begin{equation}
\mathcal{P}_{RNN}=\sum_{j=2}^{\mathcal{L}}n_{j-1}n_{j}+n_{j}^{2}+n_{j}-n_{\mathcal{L}}^{2},\label{eq:42}
\end{equation}
with $n_{1}$ denotes the number of RNN inputs at each time step. 

For the CV-TDNN {[}\ref{A.-T.-Kristensen,}{]}, using the previous
implementation assumptions for the CV additions and multiplications,
the number of FLOPs and parameters can be given by 

\vspace{-2mm}

\begin{equation}
\mathscr{\mathcal{F}_{\mathit{CV-TDNN}}}=10\left\{ \sum_{j=2}^{\mathcal{L}}n_{j-1}n_{j}\right\} +\mathcal{F^{\mathit{act,j}}}n_{j},\label{eq:43}
\end{equation}
\vspace{-2mm}

\begin{equation}
\mathcal{P}_{CV-TDNN}=2\left\{ \sum_{j=2}^{\mathcal{L}}n_{j-1}n_{j}+n_{j}\right\} .\hspace*{0.6cm}\label{eq:44}
\end{equation}

For the LWGS {[}\ref{Our paper }{]} employing $n_{i}$ CV inputs
and a single hidden layer with $n_{h}$ neurons, the number of FLOPs
and parameters can be expressed as

\begin{equation}
\mathscr{\mathcal{F}_{\mathit{LWGS}}}=10\left\{ \left(\sum_{j=1}^{\mathcal{\mathit{n_{h}}}}j\right)+\mathcal{\mathit{n_{i}}}\right\} +\mathcal{F^{\mathit{act}}}n_{h},\label{eq:45}
\end{equation}
\vspace{-2mm}

\begin{equation}
\mathcal{P}_{LWGS}=2\left\{ \left(\sum_{j=1}^{\mathcal{\mathit{n_{h}}}}j\right)+\mathcal{\mathit{n_{i}}}+\mathcal{\mathit{n_{h}}+}1\right\} .\label{eq:46}
\end{equation}
Likewise, for the MWGS {[}\ref{Our paper }{]} employing $n_{i}$
CV inputs, a $\mathcal{W}$ window size, and a single hidden layer
with $n_{h}$ neurons, the number of FLOPs and parameters can be obtained
as follows:

\vspace{-2mm}

\begin{equation}
\mathscr{\mathcal{F}_{\mathit{MWGS}}}=10\left\{ \mathcal{\mathit{n_{i}}}+\mathcal{W}\left(\mathcal{\mathit{n_{h}}}-1\right)+\mathcal{\mathit{n_{h}}}\right\} +\mathcal{F^{\mathit{act}}}n_{h},\label{eq:47}
\end{equation}

\vspace{-2mm}

\begin{equation}
\mathcal{P}_{MWGS}=2\left\{ \mathcal{\mathit{n_{i}}}+\mathcal{W}\left(\mathcal{\mathit{n_{h}}}-1\right)+2\mathcal{\mathit{n_{h}}}+1\right\} .\label{eq:48}
\end{equation}

Finally, the total number of FLOPs of any of the aforementioned NN-based
cancelers can be expressed as \vspace{-2mm}

\begin{equation}
\mathscr{\mathcal{F}_{\mathit{NN-\mathcal{C}}}}=\mathscr{\mathcal{F}_{\mathit{lin}}}+\mathscr{\mathcal{F}_{\mathit{NN}}}+2,\label{eq:49}
\end{equation}
where $\mathscr{\mathcal{F}_{\mathit{lin}}}$ represents the linear
canceler complexity, which can be calculated by (\ref{eq:19}), while
$\mathscr{\mathscr{\mathcal{F}_{\mathit{NN}}}}$ indicates the NN
complexity (e.g., $\mathscr{\mathcal{F}_{\mathit{HCRNN}}}$, $\mathscr{\mathcal{F}_{\mathit{HCRDNN}}}$,
$\mathscr{\mathcal{F}_{\mathit{RV-TDNN}}}$). The two FLOPs added
in (\ref{eq:49}) represent the number of real additions required
to sum the outputs of the linear and non-linear cancelers in (\ref{eq:8})
and (\ref{eq:9}), respectively {[}\ref{Hardware implementation of neural self-interference cancellation}{]}.

\vspace{-0.5mm}

\section{\label{sec:Training-of-the Proposed HCRNN and HCRDNN}Optimum Settings
of the Proposed HCRNN and HCRDNN Architectures}

In this section, the optimum settings for training the proposed HCRNN
and HCRDNN architectures are obtained. Specifically, the number of
convolutional filters, filter size, number of recurrent and dense
layers' neurons, activation functions in each layer, learning rate,
batch size, and training optimizer are selected in such a way that
results in a proper cancellation performance while maintaining low
computational complexity. In the following, we firstly describe the
dataset employed to train the proposed NN architectures, then the
network settings for the training process are optimized. \vspace{-2mm}

\subsection{\label{subsec:Training-Dataset}Training Dataset }

In the following experiments, the public dataset employed in {[}\ref{A.-Balatsoukas-Stimming,-"Non-li}{]}
and {[}\ref{A.-T.-Kristensen,}{]} is utilized to train and verify
the proposed NN architectures. The dataset is produced using a realistic
FD test-bed, which generates a 10 MHz quadrature phase-shift keying
modulated-orthogonal frequency division multiplexing (OFDM) signal
with an average transmit power of 10 dBm. The OFDM signal employs
1024 sub-carriers and is sampled at 20 MHz. The generated dataset
contains 20,480 time-domain baseband samples, which we split into
two distinct parts. The first part is used for training and consists
of 90\% of samples, whereas the second is employed for verification
and contains the rest 10\%. The employed hardware test-bed provides
53 dB passive analog RF cancellation using physical separation between
the transmit and receive antennas; hence, in this work, no further
active analog cancellation techniques are employed since the use of
digital cancellation after passive analog suppression is adequate
to bring the SI signal's power down to the receiver's noise level
{[}\ref{A.-Balatsoukas-Stimming,-"Non-li}{]}, {[}\ref{A.-T.-Kristensen,}{]}.
A summary of the parameters employed to generate the aforementioned
dataset is presented in Table. \ref{tab:Specifications-of-the}. 

\begin{table*}
\vspace{1mm}
{\scriptsize{}\caption{\textcolor{red}{\label{tab:Specifications-of-the} }Summary of the
parameters employed to generate the dataset utilized for training
and verifying the proposed NN architectures.}
}{\scriptsize\par}
\centering{}{\scriptsize{}}%
\begin{tabular}{>{\centering}p{5cm}>{\centering}p{5cm}}
\toprule 
\textbf{\scriptsize{}Parameter} & \textbf{\scriptsize{}Value}\tabularnewline
\midrule
{\scriptsize{}Type of modulation} & {\scriptsize{}QPSK-modulated OFDM}\tabularnewline
{\scriptsize{}Pass-band bandwidth} & {\scriptsize{}10 MHz}\tabularnewline
{\scriptsize{}Number of carriers} & {\scriptsize{}1024}\tabularnewline
{\scriptsize{}Sampling frequency} & {\scriptsize{}20 MHz}\tabularnewline
{\scriptsize{}Average transmit power} & {\scriptsize{}10 dBm}\tabularnewline
{\scriptsize{}Passive analog suppression} & {\scriptsize{}53 dB}\tabularnewline
{\scriptsize{}Active analog suppression} & {\scriptsize{}N/A}\tabularnewline
{\scriptsize{}Total analog cancellation} & {\scriptsize{}53 dB}\tabularnewline
{\scriptsize{}Dataset size} & {\scriptsize{}20,480 samples}\tabularnewline
\bottomrule
\end{tabular}{\scriptsize\par}
\end{table*}

\begin{table*}[t]
\vspace{2mm}
{\scriptsize{}\caption{\label{tab:Proposed-HCRNN-optimum}Selection of the optimum configuration
of the proposed HCRNN.}
}{\scriptsize\par}
\raggedright{}{\scriptsize{}}%
\begin{tabular}{>{\raggedright}p{0.7cm}>{\raggedright}m{0.75cm}>{\raggedright}m{0.9cm}>{\raggedright}m{0.85cm}>{\raggedright}p{0.5cm}>{\centering}m{0.7cm}>{\centering}m{0.8cm}>{\raggedright}p{0.7cm}>{\raggedright}m{0.75cm}>{\raggedright}m{0.9cm}>{\raggedright}m{0.85cm}>{\raggedright}p{0.5cm}>{\centering}m{1cm}>{\centering}m{0.8cm}}
\toprule 
\addlinespace
\multirow{2}{0.7cm}{\centering{}\textbf{\scriptsize{}Config. \#}} & \multirow{2}{0.75cm}{\centering{}\textbf{\scriptsize{}\# Filters}} & \multirow{2}{0.9cm}{\centering{}\textbf{\scriptsize{}Filter size}} & \multirow{2}{0.85cm}{\centering{}\textbf{\scriptsize{}\# Rec. neurons}} & \multirow{2}{0.5cm}{\centering{}\textbf{\scriptsize{}SI Canc. (dB)}} & \multicolumn{2}{c}{\textbf{\scriptsize{}Complexity}} & \multirow{2}{0.7cm}{\centering{}\textbf{\scriptsize{}Config. \#}} & \multirow{2}{0.75cm}{\centering{}\textbf{\scriptsize{}\# Filters}} & \multirow{2}{0.9cm}{\centering{}\textbf{\scriptsize{}Filter size}} & \multirow{2}{0.85cm}{\centering{}\textbf{\scriptsize{}\# Rec. neurons}} & \multirow{2}{0.5cm}{\centering{}\textbf{\scriptsize{}SI Canc. (dB)}} & \multicolumn{2}{c}{\textbf{\scriptsize{}Complexity}}\tabularnewline
\cmidrule{6-7} \cmidrule{7-7} \cmidrule{13-14} \cmidrule{14-14} 
\addlinespace
 &  &  &  &  & {\tiny{}\# Par.} & {\tiny{}\# FLOPs} &  &  &  &  &  & {\tiny{}\# Par.} & {\tiny{}\# FLOPs}\tabularnewline
\midrule
\centering{}{\scriptsize{}1} & \centering{}{\scriptsize{}2} & \centering{}{\scriptsize{}4$\times$1$\times$1} & \centering{}{\scriptsize{}6} & \centering{}{\scriptsize{}43.01} & \centering{}{\scriptsize{}116} & \centering{}{\scriptsize{}640} & \centering{}{\scriptsize{}17} & \centering{}{\scriptsize{}2} & \centering{}{\scriptsize{}4$\times$1$\times$1} & \centering{}{\scriptsize{}10} & \centering{}{\scriptsize{}44.18} & \centering{}{\scriptsize{}208} & \centering{}{\scriptsize{}820}\tabularnewline
\centering{}{\scriptsize{}2} & \centering{}{\scriptsize{}2} & \centering{}{\scriptsize{}8$\times$1$\times$1} & \centering{}{\scriptsize{}6} & \centering{}{\scriptsize{}42.60} & \centering{}{\scriptsize{}124} & \centering{}{\scriptsize{}688} & \centering{}{\scriptsize{}18} & \centering{}{\scriptsize{}2} & \centering{}{\scriptsize{}9$\times$1$\times$1} & \centering{}{\scriptsize{}10} & \centering{}{\scriptsize{}44.05} & \centering{}{\scriptsize{}218} & \centering{}{\scriptsize{}840}\tabularnewline
\centering{}{\scriptsize{}3} & \centering{}{\scriptsize{}2} & \centering{}{\scriptsize{}6$\times$1$\times$1} & \centering{}{\scriptsize{}7} & \centering{}{\scriptsize{}42.92} & \centering{}{\scriptsize{}140} & \centering{}{\scriptsize{}735} & \centering{}{\scriptsize{}19} & \centering{}{\scriptsize{}2} & \centering{}{\scriptsize{}10$\times$1$\times$1} & \centering{}{\scriptsize{}10} & \centering{}{\scriptsize{}44.06} & \centering{}{\scriptsize{}220} & \centering{}{\scriptsize{}796}\tabularnewline
\centering{}{\scriptsize{}4} & \centering{}{\scriptsize{}2} & \centering{}{\scriptsize{}4$\times$1$\times$1} & \centering{}{\scriptsize{}8} & \centering{}{\scriptsize{}43.75} & \centering{}{\scriptsize{}158} & \centering{}{\scriptsize{}722} & \centering{}{\scriptsize{}20} & \centering{}{\scriptsize{}2} & \centering{}{\scriptsize{}11$\times$1$\times$1} & \centering{}{\scriptsize{}10} & \centering{}{\scriptsize{}44.19} & \centering{}{\scriptsize{}222} & \centering{}{\scriptsize{}736}\tabularnewline
\centering{}{\scriptsize{}5} & \centering{}{\scriptsize{}2} & \centering{}{\scriptsize{}5$\times$1$\times$1} & \centering{}{\scriptsize{}8} & \centering{}{\scriptsize{}43.81} & \centering{}{\scriptsize{}160} & \centering{}{\scriptsize{}758} & \centering{}{\scriptsize{}21} & \centering{}{\scriptsize{}2} & \centering{}{\scriptsize{}12$\times$1$\times$1} & \centering{}{\scriptsize{}10} & \centering{}{\scriptsize{}44.31} & \centering{}{\scriptsize{}224} & \centering{}{\scriptsize{}660}\tabularnewline
\centering{}{\scriptsize{}6} & \centering{}{\scriptsize{}2} & \centering{}{\scriptsize{}8$\times$1$\times$1} & \centering{}{\scriptsize{}8} & \centering{}{\scriptsize{}43.89} & \centering{}{\scriptsize{}166} & \centering{}{\scriptsize{}770} & \centering{}{\scriptsize{}22} & \centering{}{\scriptsize{}3} & \centering{}{\scriptsize{}11$\times$1$\times$1} & \centering{}{\scriptsize{}6} & \centering{}{\scriptsize{}43.66} & \centering{}{\scriptsize{}154} & \centering{}{\scriptsize{}718}\tabularnewline
\centering{}{\scriptsize{}7} & \centering{}{\scriptsize{}2} & \centering{}{\scriptsize{}10$\times$1$\times$1} & \centering{}{\scriptsize{}8} & \centering{}{\scriptsize{}43.89} & \centering{}{\scriptsize{}170} & \centering{}{\scriptsize{}698} & \centering{}{\scriptsize{}23} & \centering{}{\scriptsize{}3} & \centering{}{\scriptsize{}12$\times$1$\times$1} & \centering{}{\scriptsize{}6} & \centering{}{\scriptsize{}43.97} & \centering{}{\scriptsize{}157} & \centering{}{\scriptsize{}604}\tabularnewline
\centering{}{\scriptsize{}8} & \centering{}{\scriptsize{}2} & \centering{}{\scriptsize{}11$\times$1$\times$1} & \centering{}{\scriptsize{}8} & \centering{}{\scriptsize{}43.68} & \centering{}{\scriptsize{}172} & \centering{}{\scriptsize{}638} & \centering{}{\scriptsize{}24} & \centering{}{\scriptsize{}3} & \centering{}{\scriptsize{}11$\times$1$\times$1} & \centering{}{\scriptsize{}7} & \centering{}{\scriptsize{}44.11} & \centering{}{\scriptsize{}176} & \centering{}{\scriptsize{}761}\tabularnewline
\centering{}{\scriptsize{}9} & \centering{}{\scriptsize{}2} & \centering{}{\scriptsize{}5$\times$1$\times$1} & \centering{}{\scriptsize{}9} & \centering{}{\scriptsize{}44.03} & \centering{}{\scriptsize{}184} & \centering{}{\scriptsize{}805} & \centering{}{\scriptsize{}25} & \centering{}{\scriptsize{}3} & \centering{}{\scriptsize{}12$\times$2$\times$1} & \centering{}{\scriptsize{}7} & \centering{}{\scriptsize{}43.54} & \centering{}{\scriptsize{}194} & \centering{}{\scriptsize{}599}\tabularnewline
\centering{}{\scriptsize{}10} & \centering{}{\scriptsize{}2} & \centering{}{\scriptsize{}6$\times$1$\times$1} & \centering{}{\scriptsize{}9} & \centering{}{\scriptsize{}44.03} & \centering{}{\scriptsize{}186} & \centering{}{\scriptsize{}825} & \centering{}{\scriptsize{}26} & \centering{}{\scriptsize{}3} & \centering{}{\scriptsize{}11$\times$1$\times$1} & \centering{}{\scriptsize{}8} & \centering{}{\scriptsize{}44.10} & \centering{}{\scriptsize{}200} & \centering{}{\scriptsize{}808}\tabularnewline
\centering{}{\scriptsize{}11} & \centering{}{\scriptsize{}2} & \centering{}{\scriptsize{}7$\times$1$\times$1} & \centering{}{\scriptsize{}9} & \centering{}{\scriptsize{}44.02} & \centering{}{\scriptsize{}188} & \centering{}{\scriptsize{}829} & \centering{}\textbf{\scriptsize{}27} & \centering{}\textbf{\scriptsize{}3} & \centering{}\textbf{\scriptsize{}12$\times$1$\times$1} & \centering{}\textbf{\scriptsize{}9} & \centering{}\textbf{\scriptsize{}44.44} & \centering{}\textbf{\scriptsize{}229} & \centering{}\textbf{\scriptsize{}745}\tabularnewline
\centering{}{\scriptsize{}12} & \centering{}{\scriptsize{}2} & \centering{}{\scriptsize{}9$\times$1$\times$1} & \centering{}{\scriptsize{}9} & \centering{}{\scriptsize{}43.93} & \centering{}{\scriptsize{}192} & \centering{}{\scriptsize{}789} & \centering{}{\scriptsize{}28} & \centering{}{\scriptsize{}3} & \centering{}{\scriptsize{}13$\times$1$\times$1} & \centering{}{\scriptsize{}9} & \centering{}{\scriptsize{}43.94} & \centering{}{\scriptsize{}232} & \centering{}{\scriptsize{}607}\tabularnewline
\centering{}{\scriptsize{}13} & \centering{}{\scriptsize{}2} & \centering{}{\scriptsize{}10$\times$1$\times$1} & \centering{}{\scriptsize{}9} & \centering{}{\scriptsize{}44.08} & \centering{}{\scriptsize{}194} & \centering{}{\scriptsize{}745} & \centering{}{\scriptsize{}29} & \centering{}{\scriptsize{}3} & \centering{}{\scriptsize{}11$\times$2$\times$1} & \centering{}{\scriptsize{}9} & \centering{}{\scriptsize{}43.85} & \centering{}{\scriptsize{}232} & \centering{}{\scriptsize{}796}\tabularnewline
\centering{}{\scriptsize{}14} & \centering{}{\scriptsize{}2} & \centering{}{\scriptsize{}12$\times$1$\times$1} & \centering{}{\scriptsize{}9} & \centering{}{\scriptsize{}44.23} & \centering{}{\scriptsize{}198} & \centering{}{\scriptsize{}609} & \centering{}{\scriptsize{}30} & \centering{}{\scriptsize{}3} & \centering{}{\scriptsize{}12$\times$2$\times$1} & \centering{}{\scriptsize{}9} & \centering{}{\scriptsize{}44.09} & \centering{}{\scriptsize{}238} & \centering{}{\scriptsize{}685}\tabularnewline
\centering{}{\scriptsize{}15} & \centering{}{\scriptsize{}2} & \centering{}{\scriptsize{}12$\times$2$\times$1} & \centering{}{\scriptsize{}9} & \centering{}{\scriptsize{}43.61} & \centering{}{\scriptsize{}204} & \centering{}{\scriptsize{}569} & \centering{}{\scriptsize{}31} & \centering{}{\scriptsize{}3} & \centering{}{\scriptsize{}12$\times$1$\times$1} & \centering{}{\scriptsize{}10} & \centering{}{\scriptsize{}44.49} & \centering{}{\scriptsize{}257} & \centering{}{\scriptsize{}800}\tabularnewline
\centering{}{\scriptsize{}16} & \centering{}{\scriptsize{}2} & \centering{}{\scriptsize{}3$\times$1$\times$1} & \centering{}{\scriptsize{}10} & \centering{}{\scriptsize{}44.06} & \centering{}{\scriptsize{}206} & \centering{}{\scriptsize{}768} & \centering{}{\scriptsize{}32} & \centering{}{\scriptsize{}3} & \centering{}{\scriptsize{}12$\times$2$\times$1} & \centering{}{\scriptsize{}10} & \centering{}{\scriptsize{}44.12} & \centering{}{\scriptsize{}299} & \centering{}{\scriptsize{}734}\tabularnewline
\bottomrule
\end{tabular}\vspace{-2mm}
\end{table*}

\begin{table*}
\vspace{3mm}
{\scriptsize{}\caption{\label{Proposed-HCRDNN-optimum}Selection of the optimum configuration
of the proposed HCRDNN.}
}{\scriptsize\par}
\raggedright{}\textbf{\scriptsize{}\hspace*{-0.1cm}}{\scriptsize{}}%
\begin{tabular}{>{\raggedright}p{0.5cm}>{\raggedright}m{0.7cm}>{\raggedright}m{0.7cm}>{\raggedright}m{0.8cm}>{\raggedright}m{0.8cm}>{\raggedright}p{0.5cm}>{\centering}m{0.5cm}>{\centering}m{0.5cm}>{\raggedright}p{0.5cm}>{\raggedright}m{0.7cm}>{\raggedright}m{0.7cm}>{\raggedright}m{0.8cm}>{\raggedright}m{0.8cm}>{\raggedright}p{0.5cm}>{\centering}m{0.5cm}>{\centering}m{0.5cm}}
\toprule 
\addlinespace
\multirow{2}{0.5cm}{\centering{}\textbf{\scriptsize{}Config. \#}} & \multirow{2}{0.7cm}{\centering{}\textbf{\scriptsize{}\# Filters}} & \multirow{2}{0.7cm}{\centering{}\textbf{\scriptsize{}Filter size}} & \multirow{2}{0.8cm}{\raggedleft{}\textbf{\scriptsize{}\# Rec. neurons}} & \multirow{2}{0.8cm}{\centering{}\textbf{\scriptsize{}\# Dense neurons}} & \multirow{2}{0.5cm}{\centering{}\textbf{\scriptsize{}SI Canc. (dB)}} & \multicolumn{2}{c}{\textbf{\scriptsize{}Complexity}} & \multirow{2}{0.5cm}{\centering{}\textbf{\scriptsize{}Config. \#}} & \multirow{2}{0.7cm}{\centering{}\textbf{\scriptsize{}\# Filters}} & \multirow{2}{0.7cm}{\centering{}\textbf{\scriptsize{}Filter size}} & \multirow{2}{0.8cm}{\raggedleft{}\textbf{\scriptsize{}\# Rec. neurons}} & \multirow{2}{0.8cm}{\centering{}\textbf{\scriptsize{}\# Dense neurons}} & \multirow{2}{0.5cm}{\centering{}\textbf{\scriptsize{}SI Canc. (dB)}} & \multicolumn{2}{c}{\textbf{\scriptsize{}Complexity}}\tabularnewline
\cmidrule{7-8} \cmidrule{8-8} \cmidrule{15-16} \cmidrule{16-16} 
\addlinespace
 &  &  &  &  &  & {\tiny{}\#}{\tiny\par}

{\tiny{}Par.} & {\tiny{}\# }{\tiny\par}

{\tiny{}FLOPs} &  &  &  &  &  &  & {\tiny{}\#}{\tiny\par}

{\tiny{}Par.} & {\tiny{}\# }{\tiny\par}

{\tiny{}FLOPs}\tabularnewline
\midrule
\centering{}{\scriptsize{}1} & \centering{}{\scriptsize{}2} & \centering{}{\scriptsize{}7$\times$1$\times$1} & \centering{}{\scriptsize{}\hspace*{3mm}5} & \centering{}{\scriptsize{}9} & \centering{}{\scriptsize{}44.10} & {\scriptsize{}166} & {\scriptsize{}789} & \centering{}{\scriptsize{}17} & \centering{}{\scriptsize{}2} & \centering{}{\scriptsize{}12$\times$1$\times$1} & \centering{}{\scriptsize{}\hspace*{3mm}8} & \centering{}{\scriptsize{}10} & \centering{}{\scriptsize{}44.42} & {\scriptsize{}268} & {\scriptsize{}740}\tabularnewline
\centering{}{\scriptsize{}2} & \centering{}{\scriptsize{}2} & \centering{}{\scriptsize{}9$\times$1$\times$1} & \centering{}{\scriptsize{}\hspace*{3mm}5} & \centering{}{\scriptsize{}10} & \centering{}{\scriptsize{}44.10} & {\scriptsize{}178} & {\scriptsize{}755} & \centering{}{\scriptsize{}18} & \centering{}{\scriptsize{}2} & \centering{}{\scriptsize{}12$\times$1$\times$1} & \centering{}{\scriptsize{}\hspace*{3mm}9} & \centering{}{\scriptsize{}7} & \centering{}{\scriptsize{}44.42} & {\scriptsize{}264} & {\scriptsize{}734}\tabularnewline
\centering{}{\scriptsize{}3} & \centering{}{\scriptsize{}2} & \centering{}{\scriptsize{}12$\times$1$\times$1} & \centering{}{\scriptsize{}\hspace*{3mm}5} & \centering{}{\scriptsize{}11} & \centering{}{\scriptsize{}44.29} & {\scriptsize{}192} & {\scriptsize{}590} & \centering{}{\scriptsize{}19} & \centering{}{\scriptsize{}2} & \centering{}{\scriptsize{}12$\times$1$\times$1} & \centering{}{\scriptsize{}\hspace*{3mm}9} & \centering{}{\scriptsize{}8} & \centering{}{\scriptsize{}44.50} & {\scriptsize{}276} & {\scriptsize{}757}\tabularnewline
\centering{}{\scriptsize{}4} & \centering{}{\scriptsize{}2} & \centering{}{\scriptsize{}9$\times$1$\times$1} & \centering{}{\scriptsize{}\hspace*{3mm}5} & \centering{}{\scriptsize{}12} & \centering{}{\scriptsize{}44.14} & {\scriptsize{}194} & {\scriptsize{}785} & \centering{}{\scriptsize{}20} & \centering{}{\scriptsize{}2} & \centering{}{\scriptsize{}12$\times$1$\times$1} & \centering{}{\scriptsize{}\hspace*{3mm}9} & \centering{}{\scriptsize{}9} & \centering{}{\scriptsize{}44.42} & {\scriptsize{}288} & {\scriptsize{}780}\tabularnewline
\centering{}{\scriptsize{}5} & \centering{}{\scriptsize{}2} & \centering{}{\scriptsize{}10$\times$1$\times$1} & \centering{}{\scriptsize{}\hspace*{3mm}5} & \centering{}{\scriptsize{}12} & \centering{}{\scriptsize{}44.18} & {\scriptsize{}196} & {\scriptsize{}741} & \centering{}{\scriptsize{}21} & \centering{}{\scriptsize{}2} & \centering{}{\scriptsize{}12$\times$1$\times$1} & \centering{}{\scriptsize{}\hspace*{3mm}10} & \centering{}{\scriptsize{}6} & \centering{}{\scriptsize{}44.46} & {\scriptsize{}282} & {\scriptsize{}770}\tabularnewline
\centering{}{\scriptsize{}6} & \centering{}{\scriptsize{}2} & \centering{}{\scriptsize{}12$\times$1$\times$1} & \centering{}{\scriptsize{}\hspace*{3mm}5} & \centering{}{\scriptsize{}12} & \centering{}{\scriptsize{}44.27} & {\scriptsize{}200} & {\scriptsize{}605} & \centering{}{\scriptsize{}22} & \centering{}{\scriptsize{}3} & \centering{}{\scriptsize{}12$\times$1$\times$1} & \centering{}{\scriptsize{}\hspace*{3mm}5} & \centering{}{\scriptsize{}6} & \centering{}{\scriptsize{}44.11} & {\scriptsize{}175} & {\scriptsize{}635}\tabularnewline
\centering{}{\scriptsize{}7} & \centering{}{\scriptsize{}2} & \centering{}{\scriptsize{}10$\times$1$\times$1} & \centering{}{\scriptsize{}\hspace*{3mm}6} & \centering{}{\scriptsize{}9} & \centering{}{\scriptsize{}44.09} & {\scriptsize{}197} & {\scriptsize{}745} & \centering{}{\scriptsize{}23} & \centering{}{\scriptsize{}3} & \centering{}{\scriptsize{}12$\times$1$\times$1} & \centering{}{\scriptsize{}\hspace*{3mm}5} & \centering{}{\scriptsize{}7} & \centering{}{\scriptsize{}44.10} & {\scriptsize{}183} & {\scriptsize{}650}\tabularnewline
\centering{}{\scriptsize{}8} & \centering{}{\scriptsize{}2} & \centering{}{\scriptsize{}12$\times$1$\times$1} & \centering{}{\scriptsize{}\hspace*{3mm}6} & \centering{}{\scriptsize{}9} & \centering{}{\scriptsize{}44.22} & {\scriptsize{}201} & {\scriptsize{}609} & \centering{}{\scriptsize{}24} & \centering{}{\scriptsize{}3} & \centering{}{\scriptsize{}12$\times$1$\times$1} & \centering{}{\scriptsize{}\hspace*{3mm}5} & \centering{}{\scriptsize{}8} & \centering{}{\scriptsize{}44.15} & {\scriptsize{}191} & {\scriptsize{}665}\tabularnewline
\centering{}{\scriptsize{}9} & \centering{}{\scriptsize{}2} & \centering{}{\scriptsize{}11$\times$1$\times$1} & \centering{}{\scriptsize{}\hspace*{3mm}6} & \centering{}{\scriptsize{}10} & \centering{}{\scriptsize{}44.11} & {\scriptsize{}208} & {\scriptsize{}702} & \centering{}{\scriptsize{}25} & \centering{}{\scriptsize{}3} & \centering{}{\scriptsize{}12$\times$1$\times$1} & \centering{}{\scriptsize{}\hspace*{3mm}5} & \centering{}{\scriptsize{}10} & \centering{}{\scriptsize{}44.24} & {\scriptsize{}207} & {\scriptsize{}695}\tabularnewline
\centering{}{\scriptsize{}10} & \centering{}{\scriptsize{}2} & \centering{}{\scriptsize{}10$\times$1$\times$1} & \centering{}{\scriptsize{}\hspace*{3mm}6} & \centering{}{\scriptsize{}11} & \centering{}{\scriptsize{}44.25} & {\scriptsize{}215} & {\scriptsize{}779} & \centering{}{\scriptsize{}26} & \centering{}{\scriptsize{}3} & \centering{}{\scriptsize{}12$\times$1$\times$1} & \centering{}{\scriptsize{}\hspace*{3mm}5} & \centering{}{\scriptsize{}11} & \centering{}{\scriptsize{}44.17} & {\scriptsize{}215} & {\scriptsize{}710}\tabularnewline
\centering{}{\scriptsize{}11} & \centering{}{\scriptsize{}2} & \centering{}{\scriptsize{}11$\times$1$\times$1} & \centering{}{\scriptsize{}\hspace*{3mm}6} & \centering{}{\scriptsize{}11} & \centering{}{\scriptsize{}44.23} & {\scriptsize{}217} & {\scriptsize{}719} & \centering{}{\scriptsize{}27} & \centering{}\textbf{\scriptsize{}3} & \centering{}\textbf{\scriptsize{}12$\times$1$\times$1} & \centering{}{\scriptsize{}\hspace*{3mm}}\textbf{\scriptsize{}5} & \centering{}\textbf{\scriptsize{}12} & \centering{}\textbf{\scriptsize{}44.41} & \textbf{\scriptsize{}223} & \textbf{\scriptsize{}725}\tabularnewline
\centering{}{\scriptsize{}12} & \centering{}{\scriptsize{}2} & \centering{}{\scriptsize{}12$\times$1$\times$1} & \centering{}{\scriptsize{}\hspace*{3mm}6} & \centering{}{\scriptsize{}11} & \centering{}{\scriptsize{}44.26} & {\scriptsize{}219} & {\scriptsize{}643} & \centering{}{\scriptsize{}28} & \centering{}{\scriptsize{}3} & \centering{}{\scriptsize{}12$\times$1$\times$1} & \centering{}{\scriptsize{}\hspace*{3mm}6} & \centering{}{\scriptsize{}6} & \centering{}{\scriptsize{}44.28} & {\scriptsize{}199} & {\scriptsize{}682}\tabularnewline
\centering{}{\scriptsize{}13} & \centering{}{\scriptsize{}2} & \centering{}{\scriptsize{}11$\times$1$\times$1} & \centering{}{\scriptsize{}\hspace*{3mm}7} & \centering{}{\scriptsize{}11} & \centering{}{\scriptsize{}44.21} & {\scriptsize{}246} & {\scriptsize{}776} & \centering{}{\scriptsize{}29} & \centering{}{\scriptsize{}3} & \centering{}{\scriptsize{}12$\times$1$\times$1} & \centering{}{\scriptsize{}\hspace*{3mm}6} & \centering{}{\scriptsize{}8} & \centering{}{\scriptsize{}44.22} & \centering{}{\scriptsize{}217} & \centering{}{\scriptsize{}716}\tabularnewline
\centering{}\textbf{\scriptsize{}14} & \centering{}\textbf{\scriptsize{}2} & \centering{}\textbf{\scriptsize{}12$\times$1$\times$1} & \centering{}{\scriptsize{}\hspace*{3mm}}\textbf{\scriptsize{}7} & \centering{}\textbf{\scriptsize{}11} & \centering{}\textbf{\scriptsize{}44.44} & \textbf{\scriptsize{}248} & \textbf{\scriptsize{}700} & \centering{}{\scriptsize{}30} & \centering{}{\scriptsize{}3} & \centering{}{\scriptsize{}12$\times$1$\times$1} & \centering{}{\scriptsize{}\hspace*{3mm}6} & \centering{}{\scriptsize{}9} & \centering{}{\scriptsize{}44.36} & {\scriptsize{}226} & {\scriptsize{}733}\tabularnewline
\centering{}{\scriptsize{}15} & \centering{}{\scriptsize{}2} & \centering{}{\scriptsize{}12$\times$1$\times$1} & \centering{}{\scriptsize{}\hspace*{3mm}7} & \centering{}{\scriptsize{}12} & \centering{}{\scriptsize{}44.44} & {\scriptsize{}258} & {\scriptsize{}719} & \centering{}{\scriptsize{}31} & \centering{}{\scriptsize{}3} & \centering{}{\scriptsize{}12$\times$1$\times$1} & \centering{}{\scriptsize{}\hspace*{3mm}6} & \centering{}{\scriptsize{}11} & \centering{}{\scriptsize{}44.40} & \centering{}{\scriptsize{}244} & \centering{}{\scriptsize{}767}\tabularnewline
\centering{}{\scriptsize{}16} & \centering{}{\scriptsize{}2} & \centering{}{\scriptsize{}12$\times$1$\times$1} & \centering{}{\scriptsize{}\hspace*{3mm}8} & \centering{}{\scriptsize{}9} & \centering{}{\scriptsize{}44.41} & {\scriptsize{}257} & {\scriptsize{}719} & \centering{}{\scriptsize{}32} & \centering{}{\scriptsize{}3} & \centering{}{\scriptsize{}12$\times$1$\times$1} & \centering{}{\scriptsize{}\hspace*{3mm}6} & \centering{}{\scriptsize{}12} & \centering{}{\scriptsize{}44.39} & \centering{}{\scriptsize{}253} & \centering{}{\scriptsize{}784}\tabularnewline
\bottomrule
\end{tabular}\vspace{-4mm}
\end{table*}

\subsection{Optimum Configuration }

The process of selecting the optimum configuration of the proposed
HCRNN (e.g., number of convolutional filters, filter size, number
of recurrent layer's neurons) is analyzed as given in Table \ref{tab:Proposed-HCRNN-optimum}.\footnote{\label{fn:foot 1}The results in Tables \ref{tab:Proposed-HCRNN-optimum}
and \ref{Proposed-HCRDNN-optimum} are obtained using the following
default settings: $ReLU$ activation function in all hidden layers,
$\textrm{5\ensuremath{\times}1\ensuremath{0^{-3}}}$ learning rate,
158 batch size, and Adam optimization algorithm over 15 random initializations.
The aforementioned settings are optimized in the next subsections.} To that end, we test two values for the convolutional filters $L\in\{2,3\}$
and change the filter size dimensions such that $R\in\{2,3,...,13\}$,
$S\in\{1,2\}$, and $Z=1$ for all filters. Moreover, we consider
$n_{hr}\in\{4,5,...,10\}$ for the number of neurons in the recurrent
layer. Since there is a large number of combinations between $L$,
$R$, $S$, $Z$, and $n_{hr}$, we only show the NN configurations
that achieve the best cancellation-complexity trade-off in Table \ref{tab:Proposed-HCRNN-optimum}.
Specifically, in this work, choosing the optimum configuration of
the HCRNN is based on the criterion of achieving a similar or comparable
cancellation performance to the polynomial canceler with $P=5$ while
maintaining a lower implementation complexity.\footnote{We note that at $P=5$, the polynomial canceler attains 44.45 dB cancellation
{[}\ref{A.-T.-Kristensen,}{]}, while from (\ref{eq:19}), (\ref{eq:20}),
(\ref{eq:21}), (\ref{eq:22}), and (\ref{eq:23}), it requires 1558
FLOPs and 312 parameters to achieve this cancellation. } Based on this, it can be observed from Table \ref{tab:Proposed-HCRNN-optimum}
that the HCRNN configuration with three convolutional filters, $12\times1\times1$
filter size, and nine neurons in the recurrent layer achieves the
target cancellation of the polynomial canceler. Besides, it attains
the best compromise between the cancellation performance and implementation
complexity compared to its other counterparts. Similarly, the selection
of the optimum configuration of the proposed HCRDNN is explored in
Table \ref{Proposed-HCRDNN-optimum}.\textsuperscript{\ref{fn:foot 1}}
In this case, we test the number of filters $L\in\{2,3\}$, filter
dimensions $R\in\{2,3,...,13\}$, $S\in\{1,2\}$, and $Z=1$, recurrent
layer's neurons $n_{hr}\in\{4,5,...,10\}$, and dense layer's neurons
$n_{hd}\in\{4,5,...,12\}$. As can be seen from Table \ref{Proposed-HCRDNN-optimum},
two configurations of the proposed HCRDNN achieve the best compromise
among the cancellation performance, number of FLOPs, and parameters.
The first configuration (i.e., HCRDNN 1) employs two convolutional
filters, $12\times1\times1$ filter size, seven neurons in the recurrent
layer, and eleven neurons in the dense layer. However, the second
(i.e., HCRDNN 2) utilizes three convolutional filters, $12\times1\times1$
filter size, five neurons in the recurrent layer, and twelve neurons
in the dense layer. It is worth mentioning here that from Tables \ref{tab:Proposed-HCRNN-optimum}
and \ref{Proposed-HCRDNN-optimum}, the filter size of $12\times1\times1$
is shown to be the best size for the optimizable convolutional filters
employed in both the HCRNN and HCRDNN architectures. Moreover, it
is worth noting that using an additional dense layer in the proposed
HCRDNN 1 and HCRDNN 2 reduces the number of neurons in the recurrent
layer from 9 to 7 and 5 neurons, respectively, compared to the HCRNN
architecture; this significantly affects the computational complexity
of the HCRDNN model.

\begin{table*}
\vspace{2mm}
{\scriptsize{}\caption{\label{tab:HCRNN Activation-function-selection}Optimum activation
functions of the proposed HCRNN.}
}{\scriptsize\par}
\centering{}{\scriptsize{}}%
\begin{tabular}{ccccc}
\toprule 
\textbf{\scriptsize{}Config. \#} & \textbf{\scriptsize{}Conv. layer} & \textbf{\scriptsize{}Rec. layer} & \textbf{\scriptsize{}SI Canc. (dB)} & \textbf{\scriptsize{}FLOPs}\tabularnewline
\midrule
{\scriptsize{}1} & {\scriptsize{}$tanh$} & {\scriptsize{}$tanh$} & {\scriptsize{}44.61} & {\scriptsize{}850}\tabularnewline
{\scriptsize{}2} & {\scriptsize{}$tanh$} & {\scriptsize{}$ReLU$} & {\scriptsize{}44.50} & {\scriptsize{}805}\tabularnewline
{\scriptsize{}3} & {\scriptsize{}$tanh$} & {\scriptsize{}$Sigm$} & {\scriptsize{}44.15} & {\scriptsize{}832}\tabularnewline
{\scriptsize{}4} & {\scriptsize{}$ReLU$} & {\scriptsize{}$tanh$} & {\scriptsize{}44.37} & {\scriptsize{}790}\tabularnewline
\textbf{\scriptsize{}5} & \textbf{\scriptsize{}$\boldsymbol{ReLU}$} & \textbf{\scriptsize{}$\boldsymbol{ReLU}$} & \textbf{\scriptsize{}44.44} & \textbf{\scriptsize{}745}\tabularnewline
{\scriptsize{}6} & {\scriptsize{}$ReLU$} & {\scriptsize{}$Sigm$} & {\scriptsize{}44.41} & {\scriptsize{}772}\tabularnewline
{\scriptsize{}7} & {\scriptsize{}$Sigm$} & {\scriptsize{}$tanh$} & {\scriptsize{}44.27} & {\scriptsize{}826}\tabularnewline
{\scriptsize{}8} & {\scriptsize{}$Sigm$} & {\scriptsize{}$ReLU$} & {\scriptsize{}43.65} & {\scriptsize{}781}\tabularnewline
{\scriptsize{}9} & {\scriptsize{}$Sigm$} & {\scriptsize{}$Sigm$} & {\scriptsize{}43.12} & {\scriptsize{}808}\tabularnewline
\bottomrule
\end{tabular}\vspace{-2mm}
\end{table*}

\begin{table*}
\begin{raggedright}
\vspace{3mm}
{\scriptsize{}\caption{\label{tab:HCRDNN Activation Selection-}Optimum activation functions
of the proposed HCRDNN.}
}{\scriptsize\par}
\par\end{raggedright}
\centering{}{\scriptsize{}}%
\begin{tabular}{cccccccc}
\toprule 
\addlinespace
\multirow{2}{*}{\textbf{\scriptsize{}Config. \#}} & \multirow{2}{*}{\textbf{\scriptsize{}Conv. layer}} & \multirow{2}{*}{\textbf{\scriptsize{}Rec. layer}} & \multirow{2}{*}{\textbf{\scriptsize{}Dense layer}} & \multicolumn{2}{c}{\textbf{\scriptsize{}SI Canc. (dB)}} & \multicolumn{2}{c}{\textbf{\scriptsize{}FLOPs}}\tabularnewline
\cmidrule{5-8} \cmidrule{6-8} \cmidrule{7-8} \cmidrule{8-8} 
\addlinespace
 &  &  &  & \textbf{\tiny{}HCRDNN 1} & \textbf{\tiny{}HCRDNN 2} & \textbf{\tiny{}HCRDNN 1} & \textbf{\tiny{}HCRDNN 2}\tabularnewline
\midrule
{\scriptsize{}1} & {\scriptsize{}$tanh$} & {\scriptsize{}$tanh$} & {\scriptsize{}$tanh$} & {\scriptsize{}44.41} & {\scriptsize{}44.33} & {\scriptsize{}830} & {\scriptsize{}870}\tabularnewline
{\scriptsize{}2} & {\scriptsize{}$tanh$} & {\scriptsize{}$tanh$} & {\scriptsize{}$ReLU$} & {\scriptsize{}44.49} & {\scriptsize{}44.57} & {\scriptsize{}775} & {\scriptsize{}810}\tabularnewline
{\scriptsize{}3} & {\scriptsize{}$tanh$} & {\scriptsize{}$tanh$} & {\scriptsize{}$Sigm$} & {\scriptsize{}44.21} & {\scriptsize{}43.98} & {\scriptsize{}808} & {\scriptsize{}846}\tabularnewline
{\scriptsize{}4} & {\scriptsize{}$tanh$} & {\scriptsize{}$ReLU$} & {\scriptsize{}$tanh$} & {\scriptsize{}44.40} & {\scriptsize{}44.30} & {\scriptsize{}795} & {\scriptsize{}845}\tabularnewline
{\scriptsize{}5} & {\scriptsize{}$tanh$} & {\scriptsize{}$ReLU$} & {\scriptsize{}$ReLU$} & {\scriptsize{}44.36} & {\scriptsize{}44.34} & {\scriptsize{}740} & {\scriptsize{}785}\tabularnewline
{\scriptsize{}6} & {\scriptsize{}$tanh$} & {\scriptsize{}$ReLU$} & {\scriptsize{}$Sigm$} & {\scriptsize{}44.14} & {\scriptsize{}43.98} & {\scriptsize{}773} & {\scriptsize{}821}\tabularnewline
{\scriptsize{}7} & {\scriptsize{}$tanh$} & {\scriptsize{}$Sigm$} & {\scriptsize{}$tanh$} & {\scriptsize{}43.98} & {\scriptsize{}44.05} & {\scriptsize{}816} & {\scriptsize{}860}\tabularnewline
{\scriptsize{}8} & {\scriptsize{}$tanh$} & {\scriptsize{}$Sigm$} & {\scriptsize{}$ReLU$} & {\scriptsize{}44.13} & {\scriptsize{}44.34} & {\scriptsize{}761} & {\scriptsize{}800}\tabularnewline
{\scriptsize{}9} & {\scriptsize{}$tanh$} & {\scriptsize{}$Sigm$} & {\scriptsize{}$Sigm$} & {\scriptsize{}43.96} & {\scriptsize{}44.09} & {\scriptsize{}794} & {\scriptsize{}836}\tabularnewline
{\scriptsize{}10} & {\scriptsize{}$ReLU$} & {\scriptsize{}$tanh$} & {\scriptsize{}$tanh$} & {\scriptsize{}44.48} & {\scriptsize{}44.24} & {\scriptsize{}790} & {\scriptsize{}810}\tabularnewline
{\scriptsize{}11} & {\scriptsize{}$ReLU$} & {\scriptsize{}$tanh$} & {\scriptsize{}$ReLU$} & {\scriptsize{}44.54} & {\scriptsize{}44.48} & {\scriptsize{}735} & {\scriptsize{}750}\tabularnewline
{\scriptsize{}12} & {\scriptsize{}$ReLU$} & {\scriptsize{}$tanh$} & {\scriptsize{}$Sigm$} & {\scriptsize{}44.17} & {\scriptsize{}43.88} & {\scriptsize{}768} & {\scriptsize{}786}\tabularnewline
{\scriptsize{}13} & {\scriptsize{}$ReLU$} & {\scriptsize{}$ReLU$} & {\scriptsize{}$tanh$} & {\scriptsize{}44.36} & {\scriptsize{}44.14} & {\scriptsize{}755} & {\scriptsize{}785}\tabularnewline
\textbf{\scriptsize{}14} & \textbf{\scriptsize{}$\boldsymbol{ReLU}$} & \textbf{\scriptsize{}$\boldsymbol{ReLU}$} & \textbf{\scriptsize{}$\boldsymbol{ReLU}$} & \textbf{\scriptsize{}44.44} & \textbf{\scriptsize{}44.41} & \textbf{\scriptsize{}700} & \textbf{\scriptsize{}725}\tabularnewline
{\scriptsize{}15} & {\scriptsize{}$ReLU$} & {\scriptsize{}$ReLU$} & {\scriptsize{}$Sigm$} & {\scriptsize{}44.24} & {\scriptsize{}43.92} & {\scriptsize{}733} & {\scriptsize{}761}\tabularnewline
{\scriptsize{}16} & {\scriptsize{}$ReLU$} & {\scriptsize{}$Sigm$} & {\scriptsize{}$tanh$} & {\scriptsize{}44.34} & {\scriptsize{}43.89} & {\scriptsize{}776} & {\scriptsize{}800}\tabularnewline
{\scriptsize{}17} & {\scriptsize{}$ReLU$} & {\scriptsize{}$Sigm$} & {\scriptsize{}$ReLU$} & {\scriptsize{}44.31} & {\scriptsize{}44.26} & {\scriptsize{}721} & {\scriptsize{}740}\tabularnewline
{\scriptsize{}18} & {\scriptsize{}$ReLU$} & {\scriptsize{}$Sigm$} & {\scriptsize{}$Sigm$} & {\scriptsize{}44.11} & {\scriptsize{}43.95} & {\scriptsize{}754} & {\scriptsize{}776}\tabularnewline
{\scriptsize{}19} & {\scriptsize{}$Sigm$} & {\scriptsize{}$tanh$} & {\scriptsize{}$tanh$} & {\scriptsize{}44.08} & {\scriptsize{}43.89} & {\scriptsize{}814} & {\scriptsize{}846}\tabularnewline
{\scriptsize{}20} & {\scriptsize{}$Sigm$} & {\scriptsize{}$tanh$} & {\scriptsize{}$ReLU$} & {\scriptsize{}44.04} & {\scriptsize{}44.07} & {\scriptsize{}759} & {\scriptsize{}786}\tabularnewline
{\scriptsize{}21} & {\scriptsize{}$Sigm$} & {\scriptsize{}$tanh$} & {\scriptsize{}$Sigm$} & {\scriptsize{}43.90} & {\scriptsize{}43.59} & {\scriptsize{}792} & {\scriptsize{}822}\tabularnewline
{\scriptsize{}22} & {\scriptsize{}$Sigm$} & {\scriptsize{}$ReLU$} & {\scriptsize{}$tanh$} & {\scriptsize{}43.46} & {\scriptsize{}42.78} & {\scriptsize{}779} & {\scriptsize{}821}\tabularnewline
{\scriptsize{}23} & {\scriptsize{}$Sigm$} & {\scriptsize{}$ReLU$} & {\scriptsize{}$ReLU$} & {\scriptsize{}43.95} & {\scriptsize{}43.63} & {\scriptsize{}724} & {\scriptsize{}761}\tabularnewline
{\scriptsize{}24} & {\scriptsize{}$Sigm$} & {\scriptsize{}$ReLU$} & {\scriptsize{}$Sigm$} & {\scriptsize{}43.84} & {\scriptsize{}43.21} & {\scriptsize{}757} & {\scriptsize{}797}\tabularnewline
{\scriptsize{}25} & {\scriptsize{}$Sigm$} & {\scriptsize{}$Sigm$} & {\scriptsize{}$tanh$} & {\scriptsize{}42.53} & {\scriptsize{}41.98} & {\scriptsize{}800} & {\scriptsize{}836}\tabularnewline
{\scriptsize{}26} & {\scriptsize{}$Sigm$} & {\scriptsize{}$Sigm$} & {\scriptsize{}$ReLU$} & {\scriptsize{}43.02} & {\scriptsize{}43.54} & {\scriptsize{}745} & {\scriptsize{}776}\tabularnewline
{\scriptsize{}27} & {\scriptsize{}$Sigm$} & {\scriptsize{}$Sigm$} & {\scriptsize{}$Sigm$} & {\scriptsize{}42.97} & {\scriptsize{}42.94} & {\scriptsize{}778} & {\scriptsize{}812}\tabularnewline
\bottomrule
\end{tabular}\vspace{-4mm}
\end{table*}

\subsection{Optimum Activation Functions}

In this subsection, we test the cancellation performance and computational
complexity of the proposed HCRNN and HCRDNN using different activation
functions in the convolutional, recurrent, and dense (if any) layers
and select the optimum combination of activation functions that results
in the best cancellation-complexity trade-off. Specifically, in Table
\ref{tab:HCRNN Activation-function-selection}, we evaluate the performance
of the optimum HCRNN configuration, obtained in the previous subsection,
using different activation functions. As seen from Table \ref{tab:HCRNN Activation-function-selection},
the HCRNN with the $ReLU$ activation function in both convolutional
and recurrent layers achieves the target cancellation of the polynomial
canceler and provides the best compromise between the cancellation
and complexity performances. It can also be inferred from Table \ref{tab:HCRNN Activation-function-selection}
that using the $tanh$ activation function in the hidden layers of
HCRNN only enhances the SI cancellation by 0.17 dB at the cost of
augmenting the FLOPs by 14\% compared to the $ReLU$ activation function.
Hence, in this work, we employ the $ReLU$ activation function for
the HCRNN as it provides the best cancellation-complexity trade-off.

Likewise, in Table \ref{tab:HCRDNN Activation Selection-}, we test
the optimum configurations of the HCRDNN using various activation
functions. As can be observed, using the $ReLU$ activation function
in the convolutional, recurrent, and dense layers attains the best
cancellation-complexity trade-off for both HCRDNN architectures. 

\subsection{Optimum Learning Rate}

The effect of varying the learning rate on the cancellation performance
of the proposed HCRNN and HCRDNN-based cancelers is analyzed in Table
\ref{tab:Selection-of-the learnining rate}. It can be inferred that
using a learning rate of $\textrm{5\ensuremath{\times}1\ensuremath{0^{-3}}}$
achieves the best cancellation performance for both the proposed HCRNN
and HCRDNN architectures compared to the other learning rates. 

\subsection{Optimum Batch Size }

Similarly, in this subsection, we test the effect of varying the batch
size on the performance of the proposed HCRNN and HCRDNN-based cancelers.
Herein, we consider many values for the batch size, and we only show
the batch sizes that result in the best cancellation performance in
Table \ref{tab:Selection-of-the batch size}. As can be observed,
employing batch size values of 62 and 158 provide the best cancellation
performance for the proposed HCRNN and HCRDNN architectures, respectively. 

\subsection{Selected Optimizer }

The cancellation performance of the proposed HCRNN and HCRDNN-based
cancelers is analyzed using different optimizers, as illustrated in
Table \ref{tab:Selection of Optimizes-of-the}. In this work, we test
the stochastic gradient descent (SGD), adaptive moment estimation
(Adam), root mean square propagation (RMSprop), adaptive delta (Adadelta),
and adaptive max-pooling (Adamax) optimizers. As seen from Table \ref{tab:Selection of Optimizes-of-the},
the Adam optimization algorithm attains the best performance for the
HCRNN and HCRDNN compared to the other optimization techniques. 

Based on the aforementioned subsections, the optimum settings for
training the proposed HCRNN and HCRDNN architectures are summarized
in Table \ref{tab:Optimum-settings-of}.

\begin{table*}
{\scriptsize{}\caption{\label{tab:Selection-of-the learnining rate} Optimum learning rate
of the proposed HCRNN and HCRDNN.}
}{\scriptsize\par}
\centering{}{\scriptsize{}}%
\begin{tabular}{cccccc}
\toprule 
\textbf{\scriptsize{}Learning rate} & {\scriptsize{}$\mathbf{1\times10^{-2}}$} & {\scriptsize{}$\mathbf{5\times10^{-3}}$} & \textbf{\scriptsize{}$\mathbf{1\times10^{-3}}$} & {\scriptsize{}$\mathbf{5\times10^{-4}}$} & {\scriptsize{}$\mathbf{1\times10^{-4}}$}\tabularnewline
\midrule
\textbf{\scriptsize{}Network} & \multicolumn{5}{c}{\textbf{\scriptsize{}SI Cancellation (dB)}}\tabularnewline
\midrule
\textbf{\scriptsize{}HCRNN} & {\scriptsize{}44.36} & \textbf{\scriptsize{}44.44} & {\scriptsize{}44.12} & {\scriptsize{}43.64} & {\scriptsize{}38.90}\tabularnewline
\textbf{\scriptsize{}HCRDNN 1} & {\scriptsize{}44.28} & \textbf{\scriptsize{}44.44} & {\scriptsize{}43.65} & {\scriptsize{}42.92} & {\scriptsize{}39.11}\tabularnewline
\textbf{\scriptsize{}HCRDNN 2} & {\scriptsize{}44.22} & \textbf{\scriptsize{}44.41} & {\scriptsize{}44.01} & {\scriptsize{}43.26} & {\scriptsize{}39.22}\tabularnewline
\bottomrule
\end{tabular}{\scriptsize\par}
\end{table*}

\begin{table*}
\begin{centering}
\vspace{1mm}
\par\end{centering}
{\scriptsize{}\caption{\label{tab:Selection-of-the batch size}Optimum batch size of the
proposed HCRNN and HCRDNN.}
}{\scriptsize\par}
\centering{}{\scriptsize{}}%
\begin{tabular}{cccccc}
\toprule 
\multirow{1}{*}{\textbf{\scriptsize{}Batch size}} & \multirow{1}{*}{\textbf{\scriptsize{}22}} & \multirow{1}{*}{\textbf{\scriptsize{}62}} & \multirow{1}{*}{\textbf{\scriptsize{}158}} & \multirow{1}{*}{\textbf{\scriptsize{}256}} & \multirow{1}{*}{\textbf{\scriptsize{}512}}\tabularnewline
\midrule
\textbf{\scriptsize{}Network} & \multicolumn{5}{c}{\textbf{\scriptsize{}SI Cancellation (dB)}}\tabularnewline
\midrule
\textbf{\scriptsize{}HCRNN} & {\scriptsize{}44.37} & \textbf{\scriptsize{}44.50} & {\scriptsize{}44.44} & {\scriptsize{}44.40} & {\scriptsize{}44.44}\tabularnewline
\textbf{\scriptsize{}HCRDNN 1} & {\scriptsize{}44.17} & {\scriptsize{}44.38} & \textbf{\scriptsize{}44.44} & {\scriptsize{}44.39} & {\scriptsize{}44.28}\tabularnewline
\textbf{\scriptsize{}HCRDNN 2} & {\scriptsize{}44.22} & {\scriptsize{}44.29} & \textbf{\scriptsize{}44.41} & {\scriptsize{}44.37} & {\scriptsize{}44.31}\tabularnewline
\bottomrule
\end{tabular}{\scriptsize\par}
\end{table*}

\begin{table*}
\vspace{2mm}
{\scriptsize{}\caption{\label{tab:Selection of Optimizes-of-the}Selected optimizer of the
proposed HCRNN and HCRDNN.}
}{\scriptsize\par}
\centering{}{\scriptsize{}}%
\begin{tabular}{cccccc}
\toprule 
\addlinespace
\multirow{1}{*}{\textbf{\scriptsize{}Optimizer}} & \multirow{1}{*}{\textbf{\scriptsize{}SGD}} & \multirow{1}{*}{\textbf{\scriptsize{}Adam}} & \multirow{1}{*}{\textbf{\scriptsize{}RMSprop}} & \multirow{1}{*}{\textbf{\scriptsize{}Adadelta}} & \multirow{1}{*}{\textbf{\scriptsize{}Adamax}}\tabularnewline
\midrule
\textbf{\scriptsize{}Network } & \multicolumn{5}{c}{\textbf{\scriptsize{}SI Cancellation (dB)}}\tabularnewline
\midrule
\textbf{\scriptsize{}HCRNN} & {\scriptsize{}41.11 } & \textbf{\scriptsize{}44.50 } & {\scriptsize{}44.23 } & {\scriptsize{}37.80} & {\scriptsize{}44.36 }\tabularnewline
\textbf{\scriptsize{}HCRDNN 1} & {\scriptsize{}38.37 } & \textbf{\scriptsize{}44.44 } & {\scriptsize{}43.97 } & {\scriptsize{}37.84 } & {\scriptsize{}43.93 }\tabularnewline
\textbf{\scriptsize{}HCRDNN 2} & {\scriptsize{}38.73 } & \textbf{\scriptsize{}44.41 } & {\scriptsize{}43.89} & {\scriptsize{}37.83 } & {\scriptsize{}43.95}\tabularnewline
\bottomrule
\end{tabular}\vspace{-2mm}
\end{table*}

\begin{table*}
\vspace{2mm}
{\scriptsize{}\caption{\label{tab:Optimum-settings-of}Optimum settings of the proposed HCRNN
and HCDRNN.}
}{\scriptsize\par}
\centering{}{\scriptsize{}}%
\begin{tabular}{>{\raggedright}m{2.5cm}>{\raggedright}m{2.3cm}>{\raggedright}m{2.3cm}>{\raggedright}m{2.3cm}}
\toprule 
\addlinespace
\multirow{1}{2.5cm}{\centering{}\textbf{\scriptsize{}Structure}} & \multirow{1}{2.3cm}{\centering{}\textbf{\scriptsize{}HCRNN}} & \multirow{1}{2.3cm}{\centering{}\textbf{\scriptsize{}HCRDNN 1}} & \multirow{1}{2.3cm}{\centering{}\textbf{\scriptsize{}HCRDNN 2}}\tabularnewline
\midrule
\centering{}\textbf{\scriptsize{}\# Filters} & \centering{}{\scriptsize{}3} & \centering{}{\scriptsize{}2} & \centering{}{\scriptsize{}3}\tabularnewline
\centering{}\textbf{\scriptsize{}Filter size} & \centering{}{\scriptsize{}12$\times$1$\times$1} & \centering{}{\scriptsize{}12$\times$1$\times$1} & \centering{}{\scriptsize{}12$\times$1$\times$1}\tabularnewline
\centering{}\textbf{\scriptsize{}\# Rec. neurons} & \centering{}{\scriptsize{}9} & \centering{}{\scriptsize{}7} & \centering{}{\scriptsize{}5}\tabularnewline
\centering{}\textbf{\scriptsize{}\# Dense neurons} & \centering{}{\scriptsize{}N/A} & \centering{}{\scriptsize{}11} & \centering{}{\scriptsize{}12}\tabularnewline
\centering{}\textbf{\scriptsize{}Conv. activation} & \centering{}{\scriptsize{}$ReLU$} & \centering{}{\scriptsize{}$ReLU$} & \centering{}{\scriptsize{}$ReLU$}\tabularnewline
\centering{}\textbf{\scriptsize{}Rec. activation} & \centering{}{\scriptsize{}$ReLU$} & \centering{}{\scriptsize{}$ReLU$} & \centering{}{\scriptsize{}$ReLU$}\tabularnewline
\centering{}\textbf{\scriptsize{}Dense activation} & \centering{}{\scriptsize{}N/A} & \centering{}{\scriptsize{}$ReLU$} & \centering{}{\scriptsize{}$ReLU$}\tabularnewline
\centering{}\textbf{\scriptsize{}Learning rate} & \centering{}{\scriptsize{}$\textrm{5\ensuremath{\times}1\ensuremath{0^{-3}}}$} & \centering{}{\scriptsize{}$\textrm{5\ensuremath{\times}1\ensuremath{0^{-3}}}$} & \centering{}{\scriptsize{}$\textrm{5\ensuremath{\times}1\ensuremath{0^{-3}}}$}\tabularnewline
\centering{}\textbf{\scriptsize{}Batch size} & \centering{}{\scriptsize{}$\textrm{62}$} & \centering{}{\scriptsize{}$\textrm{158}$} & \centering{}{\scriptsize{}$\textrm{158}$}\tabularnewline
\centering{}\textbf{\scriptsize{}Optimizer} & \centering{}{\scriptsize{}Adam} & \centering{}{\scriptsize{}Adam} & \centering{}{\scriptsize{}Adam}\tabularnewline
\bottomrule
\end{tabular}{\scriptsize\par}
\end{table*}

\begin{table*}
\vspace{0mm}
{\scriptsize{}\caption{\label{tab:Simulation-parameters-for}Optimum settings of the state-of-the-art
NN architectures.}
}{\scriptsize\par}
\centering{}{\scriptsize{}}%
\begin{tabular}{cccccc}
\toprule 
\textbf{\scriptsize{}Structure} & \textbf{\scriptsize{}\# Hidden layers neurons} & \textbf{\scriptsize{}Activation} & \textbf{\scriptsize{}Learning rate} & \textbf{\scriptsize{}Batch size} & \textbf{\scriptsize{}Opti-mizer}\tabularnewline
\midrule
{\scriptsize{}RV-TDNN {[}\ref{A.-Balatsoukas-Stimming,-"Non-li}{]}} & {\scriptsize{}18} & {\scriptsize{}$ReLU$} & {\scriptsize{}$\textrm{5\ensuremath{\times}1\ensuremath{0^{-3}}}$} & {\scriptsize{}22} & {\scriptsize{}Adam}\tabularnewline
{\scriptsize{}RNN {[}\ref{A.-T.-Kristensen,}{]}} & {\scriptsize{}20} & {\scriptsize{}$tanh$} & {\scriptsize{}$\textrm{2.5\ensuremath{\times}1\ensuremath{0^{-3}}}$} & {\scriptsize{}158} & {\scriptsize{}Adam}\tabularnewline
{\scriptsize{}CV-TDNN {[}\ref{A.-T.-Kristensen,}{]}} & {\scriptsize{}7} & {\scriptsize{}$\mathbb{C}ReLU$} & {\scriptsize{}$\textrm{4.5\ensuremath{\times}1\ensuremath{0^{-3}}}$} & {\scriptsize{}62} & {\scriptsize{}Adam}\tabularnewline
{\scriptsize{}LWGS {[}\ref{Our paper }{]}} & {\scriptsize{}9} & {\scriptsize{}$\mathbb{C}ReLU$} & {\scriptsize{}$\textrm{4.5\ensuremath{\times}1\ensuremath{0^{-3}}}$} & {\scriptsize{}62} & {\scriptsize{}Adam}\tabularnewline
{\scriptsize{}MWGS {[}\ref{Our paper }{]}} & {\scriptsize{}12} & {\scriptsize{}$\mathbb{C}ReLU$} & {\scriptsize{}$\textrm{4.5\ensuremath{\times}1\ensuremath{0^{-3}}}$} & {\scriptsize{}62} & {\scriptsize{}Adam}\tabularnewline
{\scriptsize{}Deep RV-TDNN {[}\ref{A.-T.-Kristensen,}{]}} & {\scriptsize{}(10-10-10)} & {\scriptsize{}$ReLU$} & {\scriptsize{}$\textrm{5\ensuremath{\times}1\ensuremath{0^{-3}}}$} & {\scriptsize{}22} & {\scriptsize{}Adam}\tabularnewline
{\scriptsize{}Deep RNN {[}\ref{A.-T.-Kristensen,}{]}} & {\scriptsize{}(16-16-16)} & {\scriptsize{}$tanh$} & {\scriptsize{}$\textrm{2.5\ensuremath{\times}1\ensuremath{0^{-3}}}$} & {\scriptsize{}158} & {\scriptsize{}Adam}\tabularnewline
{\scriptsize{}Deep CV-TDNN {[}\ref{A.-T.-Kristensen,}{]}} & {\scriptsize{}(4-4-4)} & {\scriptsize{}$\mathbb{C}ReLU$} & {\scriptsize{}$\textrm{4.5\ensuremath{\times}1\ensuremath{0^{-3}}}$} & {\scriptsize{}62} & {\scriptsize{}Adam}\tabularnewline
\bottomrule
\end{tabular}{\scriptsize\par}
\end{table*}

\section{\label{sec:Results-and-Discussion}Results and Discussions}

In this section, the proposed and the state-of-the-art NN-based cancelers
are assessed for canceling the SI in a realistic FD system and compared
with the polynomial-based canceler's performance. The performance
assessment includes studying the proposed NN architectures' prediction
capabilities, MSE performance, achieved SI cancellation, power spectral
density (PSD) of the modeled SI signal, computational complexity,
and memory requirements. For the sake of comprehensive evaluation,
we compare the performance of the proposed NNs with the shallow and
deep RV-TDNN, RNN, and CV-TDNN introduced in {[}\ref{A.-Balatsoukas-Stimming,-"Non-li}{]}
and {[}\ref{A.-T.-Kristensen,}{]}; further, we consider the performance
of the LWGS and MWGS investigated in {[}\ref{Our paper }{]}. In this
work, all NN architectures are implemented using 3.5.7 Python software
with 2.0.0 TensorFlow and 2.3.1 Keras versions. Moreover, the NNs
are trained using the dataset described in Section \ref{subsec:Training-Dataset}
over 15 random network initializations. For the proposed NN architectures,
we employ the optimum settings summarized in Table \ref{tab:Optimum-settings-of},
whereas for the state-of-the-art NNs, we use the optimum settings
listed in Table \ref{tab:Simulation-parameters-for}. Besides, we
employ a memory length $M=13$ for the polynomial and all NN-based
cancelers. It is worth mentioning that the optimum settings for training
the proposed and the state-of-the-art NN-based cancelers, summarized
in Tables \ref{tab:Optimum-settings-of} and \ref{tab:Simulation-parameters-for}
respectively, are chosen by applying trial and error methodology such
they achieve a similar cancellation performance to the polynomial
canceler at $P=5$ (i.e., 44.45 dB cancellation).

\vspace{-2mm}

\subsection{Prediction Capabilities of the Proposed NN Architectures }

\begin{figure*}
\begin{raggedright}
\subfloat[\label{Scatter (a)}HCRNN.]{\begin{centering}
\textbf{\scriptsize{}\hspace*{-0.1cm}\hspace*{-0.1cm}}\includegraphics[scale=0.28]{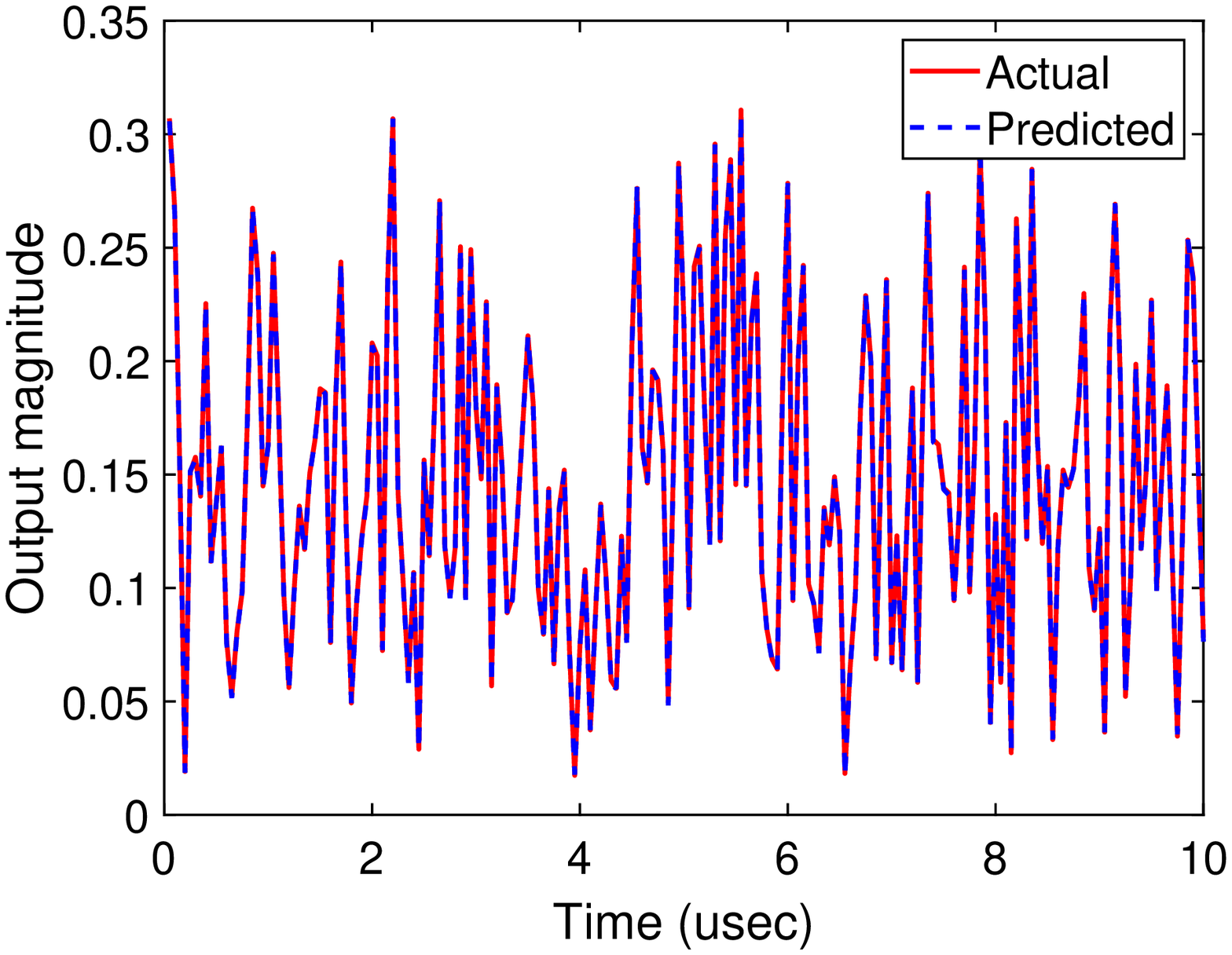}
\par\end{centering}
\centering{}} \subfloat[\label{Scatter (b)}HCRDNN 1.]{\begin{centering}
\includegraphics[scale=0.28]{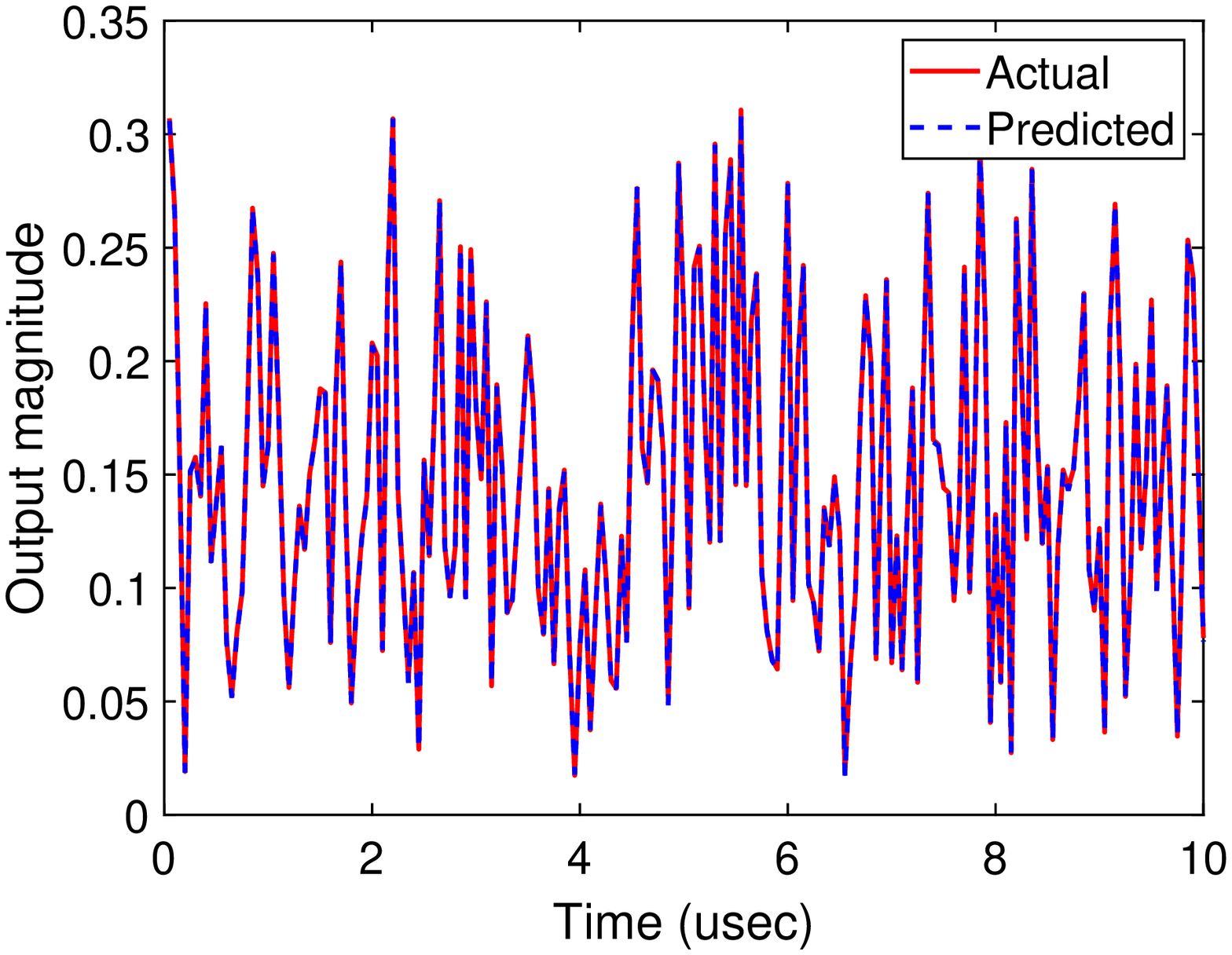}
\par\end{centering}
\centering{}} \subfloat[\label{Scatter (c)}HCRDNN 2.]{\begin{centering}
\includegraphics[scale=0.28]{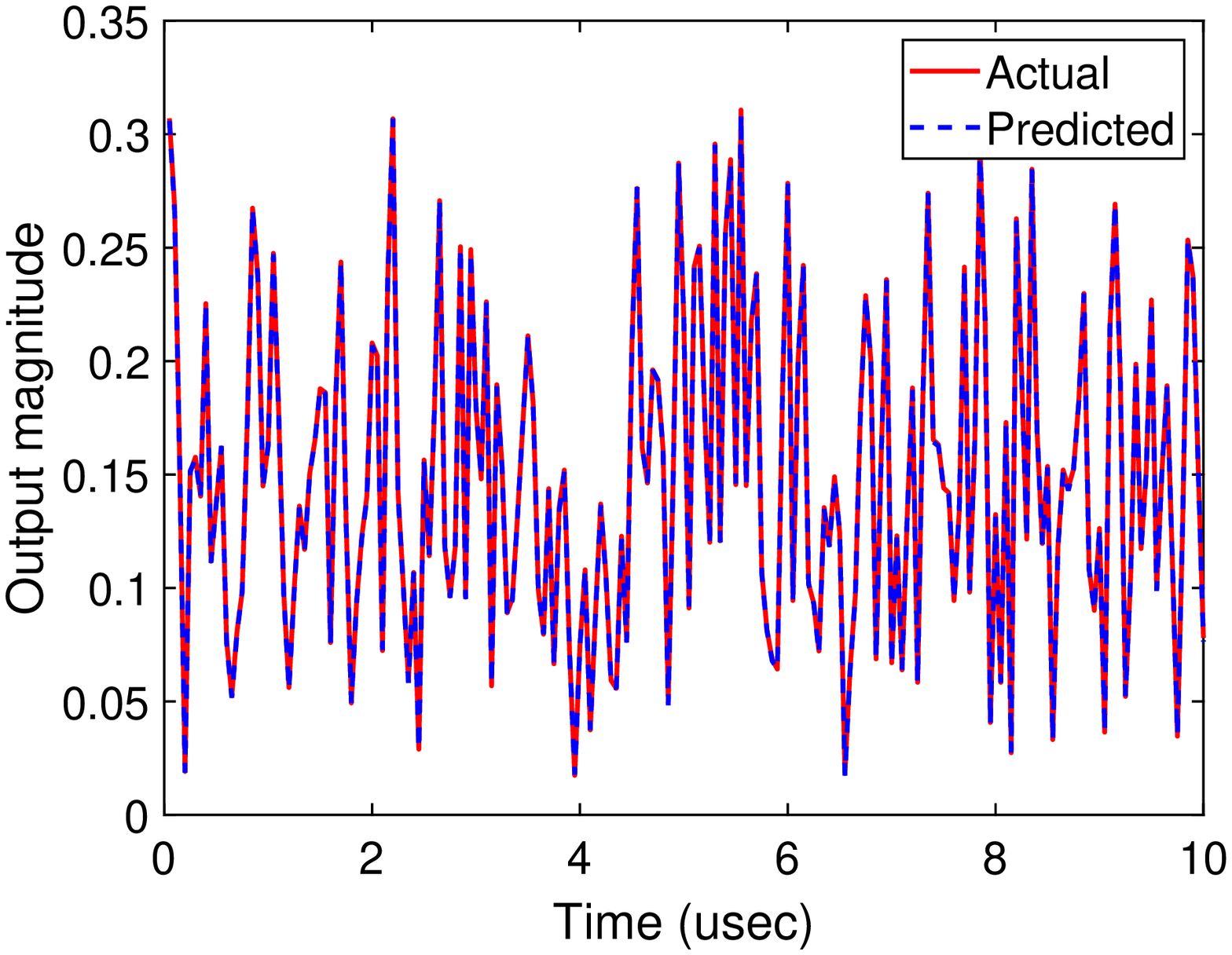}
\par\end{centering}
\centering{}}
\par\end{raggedright}
\caption{\label{fig:Actual-and-predicted}Prediction capabilities of the proposed
NN architectures.}
\end{figure*}

In this subsection, the prediction capabilities of the proposed HCRNN,
HCRDNN 1, and HCRDNN 2 architectures are assessed in Figs. \ref{fig:Actual-and-predicted}(a),
(b), and (c), respectively. Herein, we show the time-domain waveforms
for 200 input-output sample pairs of the SI signal predicted by the
proposed NN architectures and their corresponding ground-truth (actual)
values. As can be seen from the figures, there is a consistency between
the actual and predicted values of the SI signal modeled by the proposed
NNs. This consistency substantiates the ability of the proposed HCRNN
and HCRDNN-based cancelers to model the SI correctly. In addition,
it can be inferred from the figures that the proposed NNs have similar
modeling capabilities since their network settings are intentionally
set such that they achieve a comparable cancellation performance to
the polynomial-based canceler with $P=5$. 

\vspace{-2mm}

\subsection{MSE Performance }

\begin{figure*}
\begin{centering}
\includegraphics[scale=0.46]{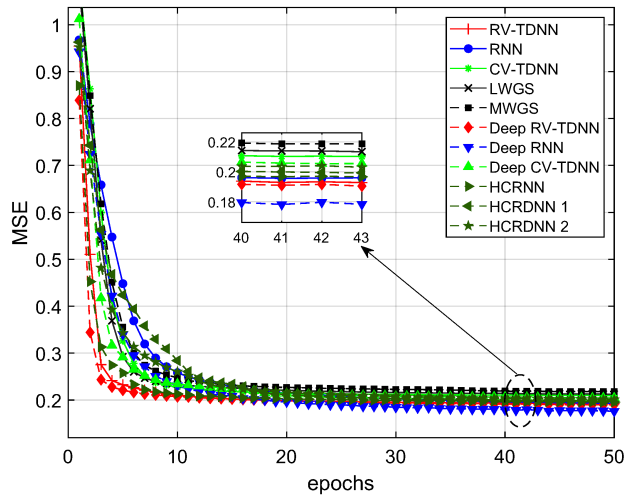}
\par\end{centering}
\centering{}\caption{\label{fig:MSE-in-training phase}MSE in the training phase for the
proposed and the state-of-the-art NN architectures.}
\end{figure*}

\begin{figure*}
\begin{centering}
\vspace{1mm}
\includegraphics[scale=0.46]{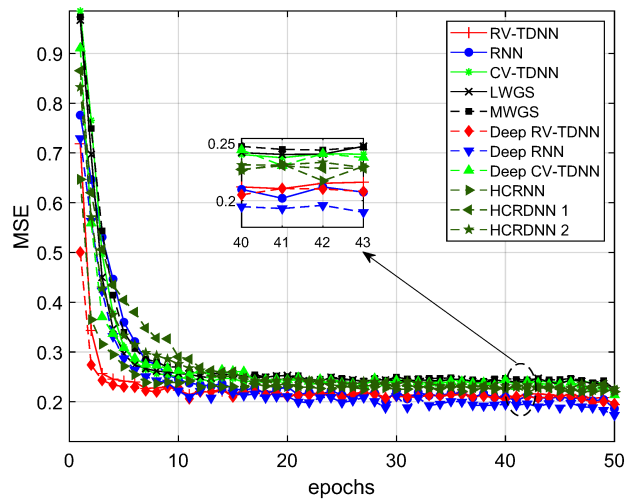}
\par\end{centering}
\centering{}\caption{\label{fig:MSE-in-testing phase }MSE in the testing phase for the
proposed and the state-of-the-art NN architectures.\vspace{-2mm}
}
\end{figure*}

The MSE performance in the training and testing phases of the proposed
NN architectures is depicted in Figs. \ref{fig:MSE-in-training phase}
and \ref{fig:MSE-in-testing phase }, respectively, compared to the
state-of-the-art NNs. The MSE is utilized to evaluate the average
squared difference between the ground-truth and predicted SI signal
modeled by the various NN architectures. As seen from the figures,
the MSE values of all NNs are comparable in both training and testing
phases (this can be clearly observed from the inset graphs in Figs.
\ref{fig:MSE-in-training phase} and \ref{fig:MSE-in-testing phase }).
As previously stated, the reason behind this is that all NNs are set
to attain a comparable cancellation performance, and that is why they
achieve a similar MSE performance. Moreover, it can be observed from
Figs. \ref{fig:MSE-in-training phase} and \ref{fig:MSE-in-testing phase }
that there are no overfitting signs for the proposed and the state-of-the-art
NNs as they perform well in both training and testing phases. Additionally,
it can be seen from the figures that the proposed NNs converge to
similar, low MSE values in both training and testing phases, which
indicates the goodness of the solution provided by the proposed NN
architectures.\vspace{-2mm}

\subsection{Achieved SI Cancellation}

\begin{figure*}
\begin{centering}
\includegraphics[scale=0.465]{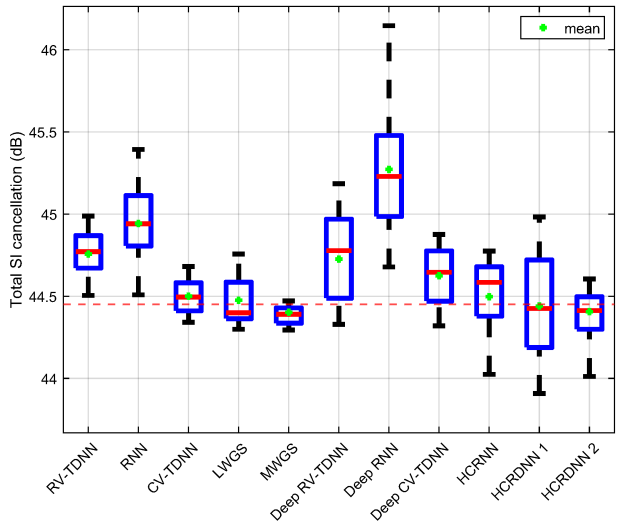}
\par\end{centering}
\centering{}\caption{\label{fig:Total-SI-cancellation of the NNs}Total SI cancellation
of the proposed and the state-of-the-art NN-based cancelers.}
\end{figure*}

The boxplots quantifying the total SI cancellation achieved by the
proposed and the state-of-the-art NN-based canceler is depicted in
Fig. \ref{fig:Total-SI-cancellation of the NNs}. Here, we calculate
the mean cancellation provided by the NN-based cancelers over the
employed 15 network initializations. As seen from the figure, all
NNs attain a comparable cancellation performance to the polynomial
canceler at $P=5$ (red dashed-line in Fig. \ref{fig:Total-SI-cancellation of the NNs}).
Specifically, the mean cancellation achieved by all NN-based cancelers
varies from 44.40 to 45.27 dB, which is very close to the polynomial
target cancellation. It is worth noting that we configure the NN settings
to achieve a comparable cancellation performance to the polynomial
canceler as it is not easy to adjust the setting to exactly achieve
the target cancellation. 

\vspace{-2mm}

\subsection{PSD Performance }

\begin{figure*}
\begin{centering}
\includegraphics[scale=0.465]{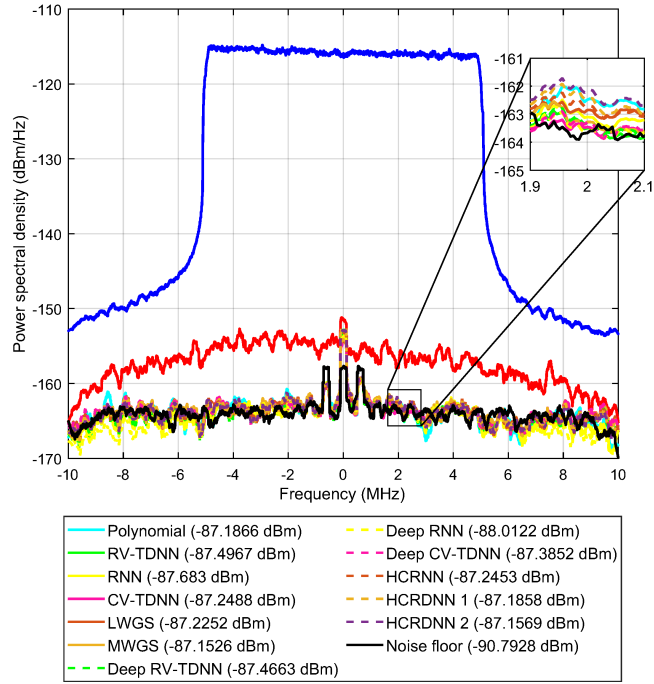}
\par\end{centering}
\centering{}\caption{\label{fig:PSD-Curves}PSD of the modeled SI signal using the proposed
and the state-of-the-art NN-based cancelers.}
\end{figure*}

In this subsection, we evaluate the PSD of the SI signal before applying
any SI cancellation techniques (blue curve in Fig. \ref{fig:PSD-Curves}).
Further, the PSD of the residual SI signal after the linear cancellation
process is assessed (red curve). Similarly, we depict the PSD of the
SI signal after performing non-linear cancellation using the polynomial
and NN-based cancelers. Finally, we show the PSD of the receiver background
noise when there is no transmission, i.e., the receiver\textquoteright s
noise floor (black curve). As can be inferred from Fig. \ref{fig:PSD-Curves},
the linear canceler provides 37.90 dB SI cancellation by bringing
the SI signal\textquoteright s power down from -42.74 dBm to -80.60
dBm. Moreover, the polynomial-based canceler attains an additional
6.6 dB cancellation by bringing the residual SI signal\textquoteright s
power down from -80.60 dBm to -87.19 dBm, making it very close to
the receiver noise floor (approximately 3 dB above the receiver\textquoteright s
noise floor). A similar task is performed by the proposed and the
state-of-the-art NN-based cancelers, in which they cancel the SI signal
after the linear canceler by 6.55-7.40 dB, making it very akin to
the receiver\textquoteright s background noise level as illustrated
in the inset graph of Fig. \ref{fig:PSD-Curves}. 

\vspace{-2mm}

\subsection{Computational Complexity and Memory Requirements }

\begin{figure*}
\begin{centering}
\includegraphics[scale=0.47]{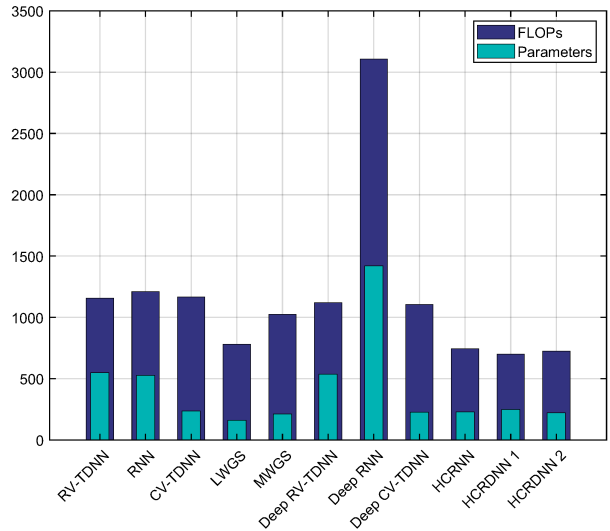}
\par\end{centering}
\centering{}\caption{\label{fig:Computational-complexity-reducti}Computational complexity
and memory requirements of the proposed and the state-of-the-art NN-based
cancelers.}
\end{figure*}

In this subsection, we assess the computational complexity of the
proposed and the state-of-the-art NN-based cancelers in terms of the
number of FLOPs and calculate the complexity reduction provided by
each canceler compared to the polynomial canceler. Similarly, we quantify
the memory requirements of each NN-based canceler in terms of the
number of parameters and calculate the amount of reduction compared
to the baseline canceler. The results of the comparison are graphically
shown in Fig. \ref{fig:Computational-complexity-reducti} and numerically
summarized in Table \ref{tab:Complexity-reduction-for-1}. 

\subsubsection{Computational Complexity}

Table \ref{tab:Complexity-reduction-for-1} illustrates the reduction
in the number of FLOPs provided by the proposed HCRNN and HCRDNN-based
cancelers compared to the polynomial and the state-of-the-art NN-based
cancelers. As seen from Table \ref{tab:Complexity-reduction-for-1},
the proposed NN-based cancelers achieve a superior enhancement in
the computational complexity compared to the polynomial-based canceler.
For instance, the HCRNN provides a 52.18\% reduction in the number
of FLOPs while maintaining the same cancellation performance of the
polynomial canceler. In addition, the HCRDNN 1 and HCRDNN 2 achieve
55.07\% and 53.47\% reduction compared to the polynomial canceler,
respectively. However, the shallow and deep RV-TDNN, RNN, and CV-TDNN
barely attain one half of the complexity reduction provided by the
proposed NN architectures. 

On the other hand, the proposed HCRNN, HCRDNN 1, and HCRDNN 2 architectures
outperform the state-of-the-art MWGS by providing 18\%, 20.92\%, and
19.32\% more reduction in FLOPs, respectively. Further, they attain
2.4\%, 5.3\%, and 3.7\% more saving in FLOPs compared to the LWGS-based
canceler, respectively. The above-mentioned results substantiate the
proposed cancelers' superiority in modeling the SI with low computational
complexity compared to the polynomial and the state-of-the-art NN-based
cancelers.

\subsubsection{Memory Requirements}

Similarly, the reduction in the number of parameters of the proposed
HCRNN and HCRDNN-based cancelers compared to the polynomial and the
state-of-the-art NN-based cancelers is illustrated in Table \ref{tab:Complexity-reduction-for-1}.
As can be observed, the shallow and deep structures of the RV-TDNN
and RNN significantly increase the required parameters compared to
the polynomial canceler. The former comes from dealing with the real
and imaginary parts separately, while the latter results from employing
a large number of feedback connections to efficiently detect the data
sequences. On the other hand, the proposed HCRNN, HCRDNN 1, and HCRDNN
2 reduce the number of parameters by 26.60\%, 20.51\%, and 28.53\%
compared to the polynomial canceler, respectively. This reduction
is due to the application of a convolutional layer before the recurrent
layer in the proposed NN architectures, which mitigates the huge memory
requirements of the recurrent layer. It can also be seen from Table
\ref{tab:Complexity-reduction-for-1} that the reduction in parameters
provided by the proposed NNs is comparable to that attained by the
shallow and deep CV-TDNN. Further, the LWGS and MWGS provide the most
reduction in the number of parameters. 

However, the proposed NNs outperform the LWGS and MWGS in the number
of FLOPs. More importantly, it can be inferred from Table \ref{tab:Complexity-reduction-for-1}
that among all NN architectures, the proposed HCRDNN 1 attains the
best reduction in FLOPs. Furthermore, the proposed HCRNN achieves
the best compromise among the cancellation performance, number of
FLOPs, and parameters. As such, the proposed NN-based cancelers offer
high design flexibility for hardware implementation, depending on
the system demands.\vspace{-2mm}

\begin{table*}
\vspace{0mm}
{\scriptsize{}\caption{\label{tab:Complexity-reduction-for-1}Reduction in the number of
FLOPs and parameters for the proposed and the state-of-the-art NN-based
cancelers compared to the polynomial canceler with $P=5$.}
}{\scriptsize\par}
\centering{}{\scriptsize{}}%
\begin{tabular}{>{\raggedright}p{3cm}>{\centering}p{1.5cm}>{\centering}p{1.5cm}>{\centering}p{1.5cm}>{\centering}p{1.5cm}>{\centering}p{1.5cm}}
\toprule 
\addlinespace
\multirow{2}{3cm}{\centering{}\textbf{\scriptsize{}Network}} & \multirow{2}{1.5cm}{\centering{}\textbf{\scriptsize{}SI Canc. (dB)}} & \multicolumn{2}{c}{\textbf{\scriptsize{}Complexity}} & \multicolumn{2}{c}{\textbf{\scriptsize{}Complexity Reduction}}\tabularnewline
\cmidrule{3-6} \cmidrule{4-6} \cmidrule{5-6} \cmidrule{6-6} 
\addlinespace
 &  & \textbf{\tiny{}\# Par.} & \textbf{\tiny{}\# FLOPs} & \textbf{\tiny{}\# Par.} & \textbf{\tiny{}\# FLOPs}\tabularnewline
\midrule
\centering{}{\scriptsize{}Polynomial ($P\negthinspace=$\,5)} & {\scriptsize{}44.45} & {\scriptsize{}312} & {\scriptsize{}1558} & {\scriptsize{}-} & {\scriptsize{}-}\tabularnewline
\centering{}{\scriptsize{}RV-TDNN} & {\scriptsize{}44.76} & {\scriptsize{}550} & {\scriptsize{}1156} & {\scriptsize{}+76.28\%} & {\scriptsize{}-25.80\%}\tabularnewline
\centering{}{\scriptsize{}RNN} & {\scriptsize{}44.94} & {\scriptsize{}528} & {\scriptsize{}1210} & {\scriptsize{}+69.23\%} & {\scriptsize{}-22.34\%}\tabularnewline
\centering{}{\scriptsize{}CV-TDNN} & {\scriptsize{}44.50} & {\scriptsize{}238} & {\scriptsize{}1166} & {\scriptsize{}-23.72\%} & {\scriptsize{}-25.16\%}\tabularnewline
\centering{}{\scriptsize{}LWGS} & {\scriptsize{}44.48} & {\scriptsize{}162} & {\scriptsize{}782} & {\scriptsize{}-48.08\%} & {\scriptsize{}-49.81\%}\tabularnewline
\centering{}{\scriptsize{}MWGS} & {\scriptsize{}44.40} & {\scriptsize{}212} & {\scriptsize{}1026} & {\scriptsize{}-32.05\%} & {\scriptsize{}-34.15\%}\tabularnewline
\centering{}{\scriptsize{}Deep RV-TDNN} & {\scriptsize{}44.73} & {\scriptsize{}538} & {\scriptsize{}1120} & {\scriptsize{}+72.44\%} & {\scriptsize{}-28.11\%}\tabularnewline
\centering{}{\scriptsize{}Deep RNN} & {\scriptsize{}45.27} & {\scriptsize{}1420} & {\scriptsize{}3106} & {\scriptsize{}+355.13\%} & {\scriptsize{}+99.36\%}\tabularnewline
\centering{}{\scriptsize{}Deep CV-TDNN} & {\scriptsize{}44.63} & {\scriptsize{}228} & {\scriptsize{}1106} & {\scriptsize{}-26.92\%} & {\scriptsize{}-29.01\%}\tabularnewline
\centering{}{\scriptsize{}HCRNN} & {\scriptsize{}44.50} & {\scriptsize{}229} & {\scriptsize{}745} & {\scriptsize{}-26.60\%} & {\scriptsize{}-52.18\%}\tabularnewline
\centering{}{\scriptsize{}HCRDNN 1} & {\scriptsize{}44.44} & {\scriptsize{}248} & {\scriptsize{}700} & {\scriptsize{}-20.51\%} & {\scriptsize{}-55.07\%}\tabularnewline
\centering{}{\scriptsize{}HCRDNN 2} & {\scriptsize{}44.41} & {\scriptsize{}223} & {\scriptsize{}725} & {\scriptsize{}-28.53\%} & {\scriptsize{}-53.47\%}\tabularnewline
\bottomrule
\end{tabular}\vspace{-5mm}
\end{table*}

\section{\textcolor{red}{\label{sec:Future-Research-Directions}}Future Research
Directions}

In this work, the proposed NN architectures have been verified using
a dataset that is captured by a single-input single-output FD test-bed.
The joint design of multiple-input multiple-output (MIMO) and FD should
be considered for B5G wireless networks {[}\ref{Beam-domain full-duplex massive MIMO}{]}-{[}\ref{A 5G-enabling technology: }{]},
and the prospects of the proposed NNs could be generalized and verified
using datasets captured by massive MIMO FD test-beds. Nevertheless,
several challenges need to be solved prior to the deployment and design
of such a system, which include but are not limited to the self- and
cross-talks among the transceiver's antennas {[}\ref{D.-Korpi,-L.}{]}.
Such challenges deserve a full investigation and will be studied in
future work. 

On the other hand, applying different machine learning techniques,
such as support vector machines (SVM) {[}\ref{SVM}{]}, for SI cancellation
in FD systems is identified as another future direction of investigation.
Several challenges related to the high computational complexity of
the SVM models can be considered in the future. \vspace{0mm}

\section{\label{sec:Conclusion}Conclusion }

In this paper, two hybrid-layers NN architectures, referred to as
hybrid-convolutional recurrent NN (HCRNN) and hybrid-convolutional
recurrent dense NN (HCRDNN), have been proposed for the first time
to model the FD system SI with reduced computational complexity. The
former exploits the weight sharing characteristics and dimensionality
reduction potential of the convolutional layer to extract the memory
effect and non-linearity incorporated into the input signal using
a reduced network scale. Moreover, it employs the high modeling capabilities
of the recurrent layer to help learn the temporal behavior of the
input signal. The latter exploits an additional dense layer to build
a deeper NN model with low complexity. The complexity analysis of
the proposed NN architectures has been conducted, and the optimum
settings for their training have been selected. Our findings demonstrate
that the proposed HCRNN and HCRDNN-based cancelers attain a reduction
in the computational complexity with 52\% and 55\% over the polynomial-based
canceler, respectively, while maintaining the same cancellation performance.
In addition, the proposed HCRNN and HCRDNN offer astounding complexity
reduction over the shallow and deep NN-based cancelers in the literature.

\vspace{-2mm}

\appendix{}

\section*{Representation of Non-linear Systems using Polynomial-based Models}

In the following, we review the common methods for representing non-linear
systems using polynomial-based models (e.g., Hammerstein and parallel-Hammerstein),
which are considered the cornerstone of approximating the SI in FD
transceivers. Herein, it is assumed that all signals are of a narrow-band
nature {[}\ref{Volterra 1}{]}. In addition, only the odd order non-linearities
are considered since the even order non-linearities lay outside the
passband {[}\ref{Volterra 1}{]} and are typically filtered by the
BPFs in the FD transceiver {[}\ref{Hardware implementation of neural self-interference cancellation}{]}.
\vspace{-2mm}

\subsection{Hammerstein Model}

The Hammerstein model is one of the widely-known models for approximating
the non-linear behavior with $M$-tap memory. In the Hammerstein model,
a static non-linearity $N(.)$ is employed in series with a linear
filter $L(.)$ in order to model the non-linearity with memory as
follows {[}\ref{Volterra 1}{]}:\vspace{-2mm}

\begin{equation}
y(n)=L\left[N\left[x\left(n\right)\right]\right]=\sum_{m=0}^{M}h_{m}\sum_{\underset{p\,odd}{p=1,}}^{P}a_{p}\,x(n-m)\left|\,x(n-m)\right|^{p-1}\hspace{-0.05cm},\label{eq:50}
\end{equation}
where $P$ indicates the order of non-linearity, whereas $h_{m}$
and $a_{p}$ represent the coefficients of the linear filter $L(.)$
and the non-linearity $N(.)$, respectively.\vspace{-2mm}

\subsection{Parallel-Hammerstein (PH) Model}

An extended version of the Hammerstein model is the PH model, which
is constructed by combining the outputs of several Hammerstein models
and can be identified as {[}\ref{Volterra 1}{]}:\vspace{-4mm}

\begin{equation}
y(n)=\sum_{\underset{p\,odd}{p=1,}}^{P}\sum_{m=0}^{M}h_{m,p}\,x(n-m)\left|x(n-m)\right|^{p-1},\label{eq:51}
\end{equation}
where $h_{m,p}$ represent the coefficients of the linear filter corresponding
to that order of non-linearity.

The PH model (\ref{eq:51}) can be rewritten as {[}\ref{D.-Korpi,-L.}{]}
as follows: \vspace{-4.5mm}

\begin{equation}
y(n)=\sum_{\underset{p\,odd}{p=1,}}^{P}\sum_{m=0}^{M}h_{m,p}\,x(n-m)^{\frac{p+1}{2}}x^{*}(n-m)^{\frac{p-1}{2}},\label{eq:52}
\end{equation}
where $\left(.\right)^{*}$ indicates the complex conjugate operator.
The PH model (\ref{eq:52}) provides a more general polynomial representation
for approximating the non-linearity with memory and is considered
the cornerstone of modeling the SI in FD transceivers.

\end{document}